\documentclass[11pt, a4paper, singlecolumn]{article}

\usepackage{color}

\usepackage[utf8]{inputenc}
\usepackage[english]{babel}

\usepackage[top=2cm, bottom=2cm, left=1.9cm, right=1.9cm]{geometry}

\usepackage{titlesec}
\titlespacing{\section}{0pt}{\parskip}{\parskip}
\titlespacing{\subsection}{0pt}{\parskip}{\parskip}
\titlespacing{\subsubsection}{0pt}{\parskip}{\parskip}

\usepackage{indentfirst}
\usepackage[export]{adjustbox}
\usepackage{graphicx}
\usepackage{amssymb}
\usepackage{amsmath}

\usepackage[dvipsnames]{xcolor}

\usepackage[figurename=Fig.]{caption}
\usepackage{caption}
\usepackage[colorlinks, citecolor=ForestGreen]{hyperref}
\usepackage{color, soul}
\usepackage{cite}

\usepackage{abstract}
\usepackage{multirow}
\usepackage[normalem]{ulem}
\usepackage{array}
\newcolumntype{*}{>{\global\let\currentrowstyle\relax}}
\newcolumntype{^}{>{\currentrowstyle}}

\newcommand\correspondingauthor{\thanks{Corresponding author.}}

\title{\textbf{Iterative Facial Image Inpainting Based on an Encoder-Generator Architecture}} 
\date{\vspace*{2pt}}
\usepackage{authblk}
\author[1]{Yahya Dogan\correspondingauthor}
\author[1, 2]{Hacer Yalim Keles}

\affil[1]{\footnotesize Ankara University, Computer Engineering Department, Turkey}
\affil[2]{\footnotesize Hacettepe University, Computer Engineering Department, Turkey}
\affil[ ]{\footnotesize yahyadogan@ankara.edu.tr, hacerkeles@cs.hacettepe.edu.tr}

\begin{document}
\maketitle

\begin{abstract}{
	\vspace*{-1.5em}
	\it Facial image inpainting is a challenging problem as it requires generating new pixels that include semantic information for masked key components in a face, e.g., eyes and nose. Recently, remarkable methods have been proposed in this field. Most of these approaches use encoder-decoder architectures and have different limitations such as allowing unique results for a given image and a particular mask. Alternatively, some optimization-based approaches generate promising results using different masks with generator networks. However, these approaches are computationally more expensive. In this paper, we propose an efficient solution to the facial image painting problem using the Cyclic Reverse Generator (CRG) architecture, which provides an encoder-generator model. We use the encoder to embed a given image to the generator space and incrementally inpaint the masked regions until a plausible image is generated; we trained a discriminator model to assess the quality of the generated images during the iterations and determine the convergence. After the generation process, for the post-processing, we utilize a Unet model that we trained specifically for this task to remedy the artifacts close to the mask boundaries. We empirically observed that even in the absence of important facial features, the encoder model is capable of embedding images in semantically rich regions in the latent space, utilizing the surrounding context in the images. Cultivating the feedback loop between the encoder and generator gradually improves the missing content in the images in an iterative fashion and only a few iterations are sufficient to generate realistic content. Since the models are not trained for particular mask types, our method allows applying sketch-based inpaintings, using a variety of mask types, and producing multiple and diverse results. We compared our method with the state-of-the-art models both quantitatively and qualitatively, and observed that our method can compete with the other models in all mask types; it is particularly better in images where larger masks are utilized.  Our code, dataset and models are available at: \href{https://github.com/yahyadogan72/iterative_facial_image_inpainting}{https://github.com/yahyadogan72/iterative\_facial\_image\_inpainting.} 
}\end{abstract}

\section{Introduction}

\begin{figure}[!h]
	\centering
	\includegraphics[scale=.4]{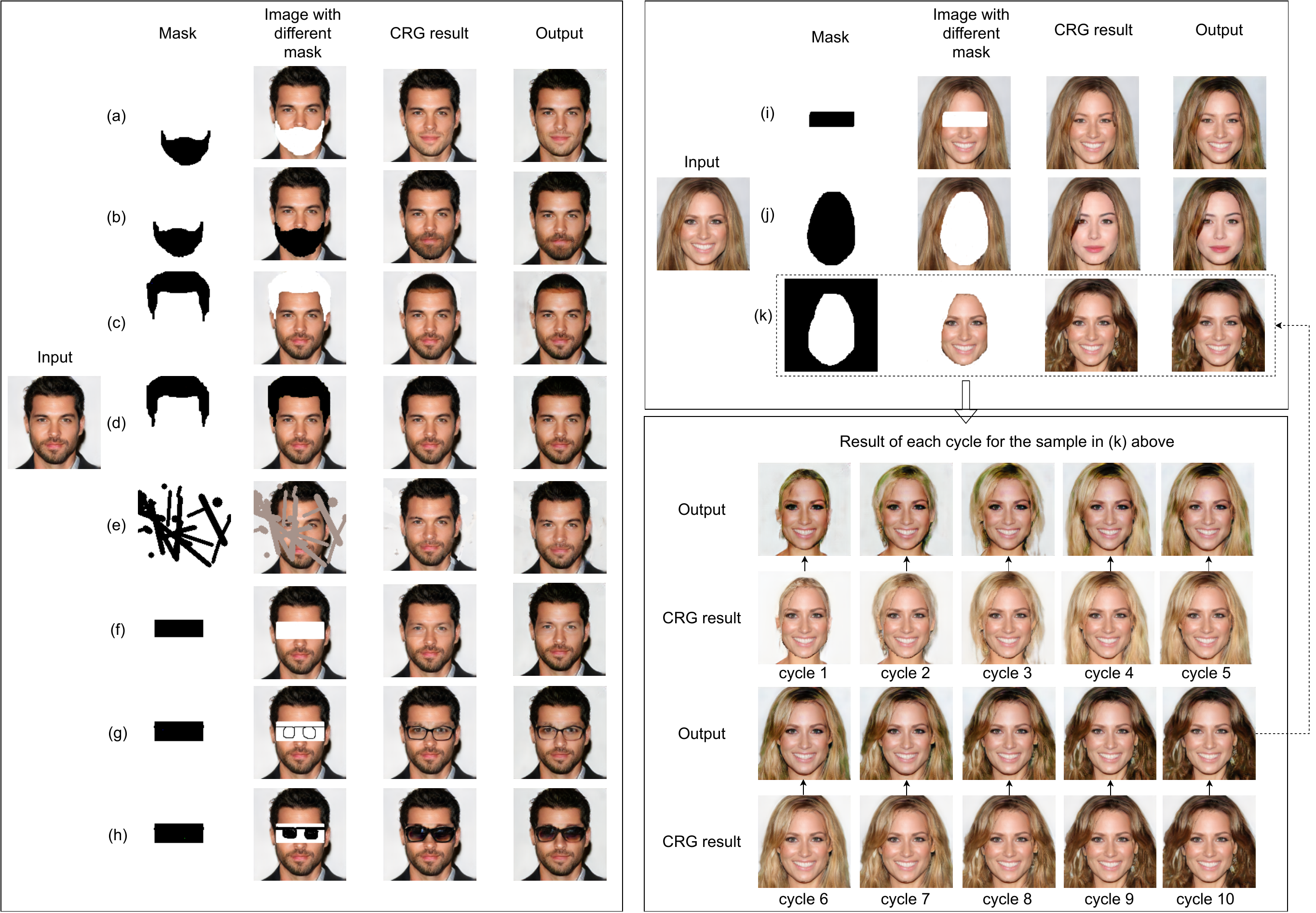}
	\centering
	\caption{Left to Right, on both columns: Input image, different mask types (regular, irregular or hand-drawn), masked image, course CRG result, fine result after post-processing. Right-bottom block shows the results of each cycle for the sample in (k) above.}
	\label{fig:different_mask_different_pixel_value}
\end{figure}

The purpose of image inpainting is to fill missing or masked regions with perceptually plausible content. This is useful for repairing damaged images, removing distracting objects/parts, or synthesizing alternative contents in masked regions. It is a long-standing topic but has recently been addressed by many researchers thanks to the advances in deep learning-based techniques, especially the generative models. Image inpainting models can be divided into two main groups; non-semantic and semantic methods. The first group, which is also referred to as the traditional approach, searches for patches from known regions of the same image or a database of images and replaces masked region with similar content that maximizes the patch similarity \cite{simakov2008summarizing, barnes2009patchmatch, darabi2012imageMelding, huang2014image}. These approaches produce texture-consistent results in completing the background inpainting problem (e.g., sky, sea, and grass), but often fail to complete complex scenes that require particular objects/parts, e.g., faces and missing parts of objects; it is hard to find patches that match exactly with a particular context. The second group, which is named as learning-based image inpainting methods can capture high-level semantics such as position, shape, color, and texture of objects, and fill masked regions with coherent structures using learned data distribution \cite{pathak2016context, li2017faceComplition, yu2018contextual, lizuka2017globally, wang2019laplacian, nazeri2019edgeconnect, wang2018gmccn, zheng2019pluralistic, hong2019deepFusion}. Facial image inpainting approaches are in this group; they require semantic content generation. For instance, if the missing part of an image is in the nose region, it is difficult to synthesize a nose fitting naturally to a face using image patches from other facial parts, if not impossible. Similar to image inpainting, facial image inpainting is useful for many tasks related to face editing, such as repairing missing regions due to damage and aging, removing particular objects/accessories from facial regions, generating different face images via masks for data augmentation, to name a few.

More recently, deep learning-based approaches show remarkable success in the task of filling in visually realistic content for masked regions. However, different approaches have different limitations or constraints in region filling: (1) Many methods are indifferent to the mask content, i.e. the sketch or pixel colors in the mask. These methods either ignore the assigned pixel value to the masked region or use the same pixel value similar to the training phase \cite{liu2018partial} \cite{yu2018contextual}. Therefore, the output remains the same even though the content of the masked region is different. (2) Some methods use rectangular masks and may not perform well for irregular shaped masks \cite{lizuka2017globally} \cite{yu2018contextual}, (3) Some models provide multiple and diverse results for masked regions, yet these models may require sketch/edge \cite{yu2019gate,nazeri2019edgeconnect, shao2020edgeColorFusion} or labels  \cite{chen2018multipleAndDiverse} as a condition during model training.

In the vast literature, most of the facial image inpainting models are based on encoder-decoder architecture; these networks learn how to fill the mask regions in a supervised manner end-to-end. The mapping is unique and constant for a particular image and the mask. There are recent approaches that show the utilization of a generator network for different promising results with varying masks \cite{yeh2017semantic, abdal2020image2stylegan++}. Our approach is similar to these works in that we also utilize a generator network for filling the content of the mask based on the surrounding context of the face image. Differently from them, however, we use an encoder to efficiently embed a given image to generator space instead of iterative latent vector optimization. This is very advantageous, since we incrementally generate plausible image after only a few iterations, i.e. 10 on the average, while optimization-based embedding requires usually a few thousand of iterations. In this context, we utilize the CRG architecture that we introduced in our preliminary work \cite{dogan2020semi} to embed a given image to a generator latent space using the encoder network. Our solution to image inpainting problem depends on the observation that the encoder embeds a given image to a  latent point in the generator space using the context of the image, even some parts of the image is masked. Moreover, the content of the mask directly influences the embedded point, hence the generated image changes accordingly. Although the generated images in the mask region contain distorted, blurry version of some face features in the early steps, when we use the generated content iteratively, the face features gradually improve in each iteration. This incremental approach generates a variety of alternative fillings in the mask regions depending on the initial mask content; since each time the changed image content results in an embedding to a different latent point. As a result, iteratively embedding to a better latent code and generating a better image in the generator space enables us at some point to reach a plausible image content. Our heuristic works successfully with the help of a discriminator, which is used to assess the quality of the generated images in each iteration.

The proposed method in this work has a two-stage process. In the first stage, the CRG models and a discriminator is used to generate a plausible image for a masked input image. In the second stage, inspired from \cite{isola2017image, liu2018partial}, a Unet model is trained to remedy the visual artifacts within the mask such as the artifacts around the boundary pixels, skin color variations, etc. We preferred using a deep network in the second stage to avoid hand-coded post-processing algorithms, e.g., Poisson image blending \cite{perez2003poisson}, fast marching methods \cite{telea2004image}.

The main contributions of this work can be summarized as follows:

\begin{itemize}
\item We propose an iterative facial image inpainting method that allows regular/irregular/hand-drawn masks as input. Our method generates inpainted images only after a few iterations.

\item Our method provides different results depending on the content, i.e. pixel values, of a masked region. To the best of our knowledge, we are the first to demonstrate this.

\item Contrary to other encoder-decoder models, our method enables image generation from a generator latent space and allows modifying undesirable regions in the generated images.

\item Our model provides sketch-based inpainting, which allows multiple and diverse results for the same masked region, without requiring conditional training.


\item Our model provides realistic images with the most plausible visual quality, with 0.83 discriminator score, compared to the state-of-the-art models, especially for large masks that cover all facial landmarks.

\end{itemize}
The rest of the paper is organized as follows. A brief summary of the related previous works is summarized in Section \ref{sec:RelatedWorks}. The proposed method is discussed in detail in Section \ref{sec:TheMethod}. The experimental results are provided in Section \ref{sec:experiments_and_results}. The paper is concluded with future directions.

\section{Related Works}
\label{sec:RelatedWorks}

Thanks to the advances in the deep learning models, recent works in the facial image inpainting domain provide plausible semantic content in images. Most of the recent works in the domain focus on designing a better encoder-decoder architecture to fill the missing parts in images. The encoder embeds an image with a missing region to a low-dimensional latent feature space, while the decoder takes the embedded latent code and reconstructs the original image. Autoencoder models are usually trained with pixel-wise reconstruction loss, which commonly causes blurry images.

Recently proposed models use a joint objective function that combines reconstruction loss with the adversarial loss that is used in GAN training \cite{goodfellow2014generative}. The adversarial loss is successfully used for high-fidelity natural image generation \cite{karras2019style}\cite{karras2020analyzing}, image to image translation \cite{richardson2021encoding}\cite{nie2021urca}, image super-resolution \cite{chan2021glean}\cite{liu2021perception}, and more \cite{gayon2020pores}\cite{ChuiPredictingStudents}\cite{wang2021small}. Context Encoder (CE) is one of the earliest works that introduce adversarial loss into the inpainting problem. CE repairs large holes using a deep convolutional neural network via learning semantic priors and meaningful hidden representations. The CE approach can produce new semantic content in the masked regions, but still it can produce inconsistent content close to the boundaries and is limited to work with fixed-size masks. Iizuka et al. \cite{lizuka2017globally} extend the CE by introducing both global and local discriminators in the adversarial loss computations, which improve visual quality significantly. Moreover, their method supports different size masks since they use a fully convolutional network. Ji and Yang \cite{ji2020image} claim that image distortion and blurring may occur when the masked regions are large, and discoloration may occur when the masked regions are placed at the edges of the image. Li et al. \cite{li2017faceComplition} used a pre-trained face parsing network to compute another loss, in addition to the global, local, and pixel-wise reconstruction losses, to enforce a more plausible and consistent result. This approach handles aligned faces well but may require post-processing methods (such as Poisson blending \cite{perez2003poisson}) to provide color consistency. Yu et al. \cite{yu2018contextual} claim that CNN-based models are ineffective in terms of modeling long-term correlations between distant contextual information and a corrupted region. This may lead to various structural artifacts between a corrupted region and its surrounding areas. They proposed a novel contextual attention layer to capture the relationship between feature patches at distant spatial locations. Recently, Wang et al. \cite{wang2019laplacian} stated that this mechanism is extremely addictive to surrounding information and may not remove all artifacts in a missing region of an image. To overcome this problem, they designed a deep network consisting of 3 parts, i.e. representation learning, feature reconstruction, and residual learning. The first part aims to capture image content and compress it into a latent feature representation. The second part has a Laplacian pyramid to recover missing regions of a face image in a coarse-to-fine fashion, progressively. The third part is a convolutional residual learning architecture that is used to eliminate color discrepancies between the corrupted area and its surroundings. Their results are realistic and natural-looking regardless of the mask locations.
 
The aforementioned models are designed to work with rectangular masks. In real-life scenarios, when inpainting in a slightly corrupted image with irregular patterns, there is a need for more flexible mask definitions. Recently, several methods have been proposed to support free-form masks to handle irregular image inpainting. Liu et al. \cite{liu2018partial} introduced a partial convolution layer where the convolution is masked and renormalized using only the valid pixels; they then apply a mask-update. Their model produces good results for irregular masks, yet their method ignores the pixels in the masked region causing different limitations, e.g. not supporting user-guided or sketch-based inpainting, on region filling. Yu et al. \cite{yu2019gate} introduced gated convolution to generalize partial convolution by learning a dynamic feature gating mechanism and also added an extension to allow user-guided inpainting. Their model produces higher quality images and more flexible results than previous methods, yet may suffer from a lack of relationship between mask and background regions, such as the symmetry of the eyes \cite{shin2020pepsi++}. Nazeri et al. \cite{nazeri2019edgeconnect} developed a new approach, which is named  Edge Connect, to provide hand-drawn and sketch-based inpainting using standard convolution operation. Their model consists of an edge generator and an image completion network, both followed by an adversarial network. The edge generator aims to detect the edges of an unknown region of an image; image completion fills the missing region in the image using detected edges as a condition. The Edge Connect method recovers images with good semantic structural consistency, but an edge artifact may occur at the boundary of completion. To handle this problem,  Hong et al. \cite{hong2019deepFusion} proposed a smooth transition named Deep Fusion Network (DFNet). They embed a few fusion blocks into different decoder layers in the Unet model to generate a flexible alpha composition map; it is similar to the hole mask but has smoother weights, especially on the boundary region.
 
 The mentioned models for image inpainting can produce plausible results for regular, irregular, or hand-drawn masks, but these models produce only one result for a single masked input. Recently, some image inpainting methods \cite{chen2018multipleAndDiverse, zheng2019pluralistic} generate multiple and diverse plausible solutions for a single mask. Chen et al. \cite{chen2018multipleAndDiverse} proposed a completely end-to-end progressive GAN model \cite{karras2017progressive} to repair face images in high-resolution, and they designed a conditional version to provide multiple results with controllable attributes, e.g., gender, smiling, etc., under arbitrary masks.  Zheng et al. \cite{zheng2019pluralistic} presented a probabilistic framework with two parallel paths to generate diverse solutions for image completion, and introduce a short+long term attention layer that exploits distant relations among decoder and encoder features to improve realism. These approaches allow producing different results that are useful in facial image inpainting, e.g., if the masked region is a mouth, both smiling and non-smiling results can be produced.

 There are very few works \cite{yeh2017semantic, abdal2020image2stylegan++} that allow both image generation and filling a masked region in an image. In Yeh et al. \cite{yeh2017semantic}'s work, semantic inpainting is considered as a constrained image generation problem. Utilizing a GAN model \cite{goodfellow2014generative}, they search for encoding of the masked image that is closest to the image in the latent space. In \cite{yeh2017semantic}, an optimization-based method is proposed to find the image closest to the original image using the rest of the image, i.e. masked part of the image is not included in the search. Abdal et al. \cite{abdal2020image2stylegan++} proposed an embedding algorithm into the $W+$ space of StyleGAN \cite{karras2019style} instead of the initial $Z$ latent space. Their method provides multiple image editing and manipulation applications, such as image inpainting, and makes the optimization-based approach interesting. However, these methods require many iterations (e.g., 1500) and a different set of hyper-parameters for each image to achieve good results. 

 As summarized above, different inpainting models have different limitations. Some of the existing methods (1) work only with fixed-size masks, (2) can produce inconsistent content between the masked region and the rest of the image, (3) handle only aligned faces well, (4) are designed to work with only rectangular masks, (5) ignore the pixel content in the masked regions, hence do not support guidance like user sketches etc., (6) produce only one (fixed) result for a single masked input, (7) require many iterations for each image to generate reasonable result. Although our model has its own limitations, it is comparably more flexible to overcome most of the listed limitations than many existing approaches.

In our work,  we utilize the CRG framework that we developed in our preliminary work \cite{dogan2020semi} to provide an efficient solution to the facial image inpainting problem via an encoder-generator architecture. Unlike the other works that have a generator network, we proposed an iterative algorithm to fill a masked region semantically with very few iterations using our encoder. Our method is flexible; it allows applying sketch-based inpaintings, using different mask types, and producing multiple and diverse results depending on an assigned pixel value to a masked region.
\section{The Method}
\label{sec:TheMethod}

In this section, we introduce our two-stage coarse-to-fine approach for facial image inpainting. The first stage makes an initial coarse prediction using our preliminary encoder-generator architecture, i.e. the CRG \cite{dogan2020semi}. In this part, we establish a cycle between the encoder and the generator and score the generated image at the end of each cycle using a discriminator network that we designed and trained specifically for this task (Section \ref{sec:Discriminator_network}). The CRG cycle is run until a plausible image content is generated (usually less than 10 iterations), and finally, the image with the highest discriminator score is selected as the coarse prediction. In general, the score is low in the first cycle, the score increases as the cycle progresses, and after a certain number of iterations, the score generated by the discriminator starts decreasing. We assume convergence before this point. 

\begin{figure}[!h]
	\centering
	\includegraphics[scale=.8]{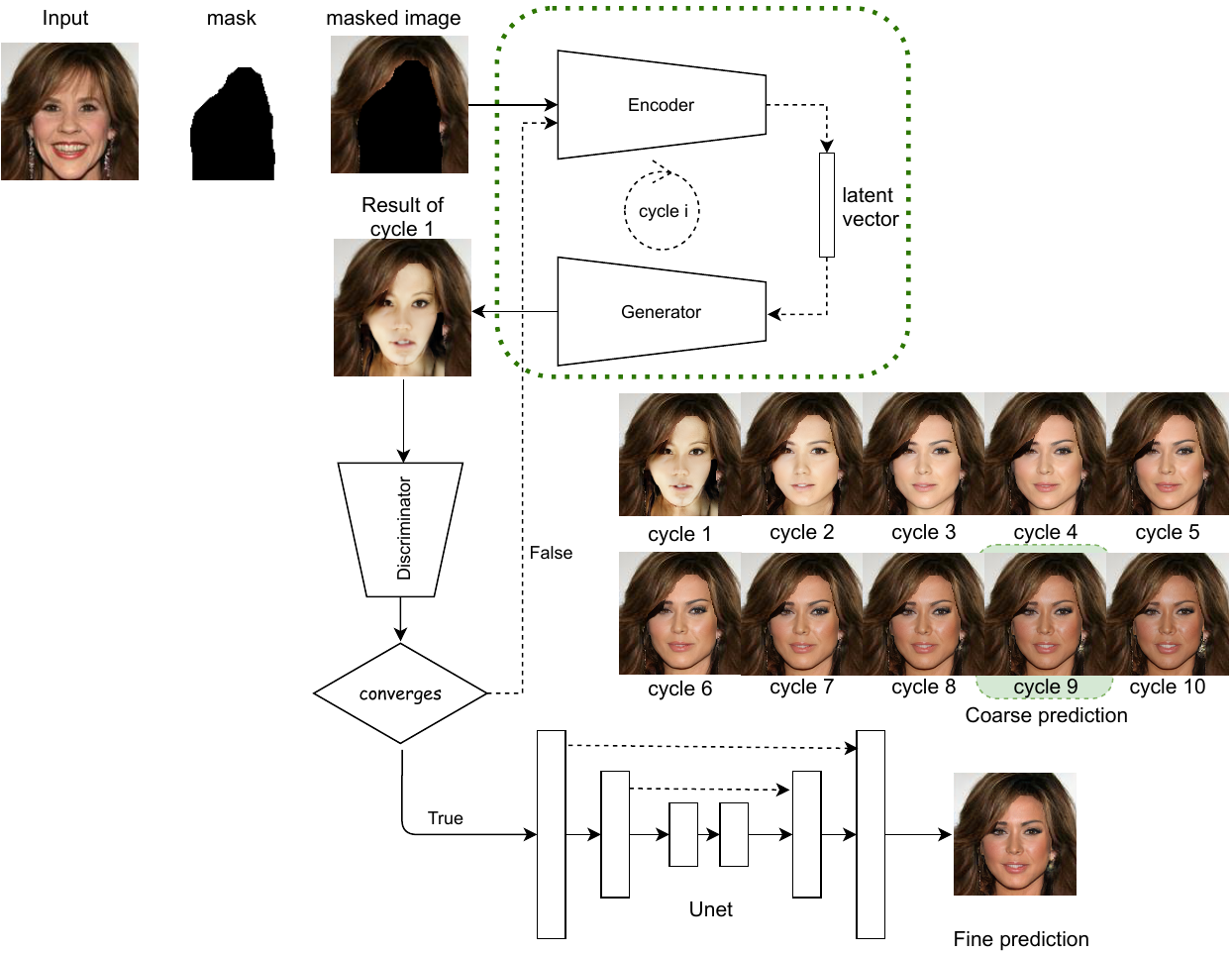}
	\centering
	\caption{Flow diagram of our proposed method.}
	\label{fig:proposed_model}
\end{figure}

The second stage of our approach can be considered as a post-processing stage. Our aim is to improve and eliminate some minor artifacts of the resultant coarse image in and around the boundaries of the masked regions. The second stage takes the generated coarse image as input and predicts a refined result using a Unet type deep network (Section \ref{sec:unet_arch}). The aim of using a network in this stage is to avoid handcrafted post-processing operations. We trained a Unet model that we adapted from \cite{isola2017image,liu2018partial} from scratch. The overall design of the proposed model is depicted in Fig. \ref{fig:proposed_model}.

We provide the flow diagram of our proposed method with an example input-mask pair in Fig. \ref{fig:proposed_model}. For a given input image and a mask, the masked image, which will be the initial input for the CRG encoder, is constructed. At the end of each cycle, the pixel values corresponding to the masked region in the original image are replaced with the corresponding pixels in the generated image in that cycle. As the encoder-generator cycle progresses, depending on the capacity of the generator, the content of the mask region is filled and iteratively improved. Depending on the discriminator scores, a converged image is determined, e.g. cycle 9. There may be small variations in the pose or color in the converged images, which may result in noticeable boundary artifacts when the masked region pixels are copied into the original image; we use our Unet model to correct such artifacts as a post-processing. The details of each step in our method are explained in the upcoming sub-sections.

\subsection{Utilization of the CRG Architecture}

In our preliminary work \cite{dogan2020semi}, we proposed the CRG architecture to find an accurate latent vector representation of an image in the generator space. CRG is composed of an encoder and a generator network; the encoder maps the image to a lower dimension space, i.e. latent vector representation, and the generator maps a given latent code to image space. This allows the reconstruction of both generated and real images. In \cite{dogan2020semi}, we used CRG for editing desired face attributes using two reference images for computing-related attribute directions in the latent space. In this work, we use the CRG encoder-generator models in a novel iterative framework for a coarse prediction of the masked regions in face images. The coarse prediction works as follows: First, the masked image is given to the CRG encoder network to compute its latent vector representation. Since masked image contains some missing regions in the masked area, the encoder maps it to a point in the latent space utilizing the context in the unmasked regions; the mask content also provides additional context. Its latent code is initially away from the one that belongs to the original unmasked image in the generator latent space. Therefore, the generated image as a result of the first cycle only contains a very coarse content in the masked regions and looks quite different from the original one. Still, it's a step further towards a realistic-looking image than the initial mask. Hence, we copy the estimated mask content to the initial image to initiate another cycle. Formally, we use (\ref{equ:crg_cycle}) to prepare the input for the next cycle.

\begin{equation}
\label{equ:crg_cycle}
\begin{split}
{I_{0} = M\odot I_{org}}\\
{I_{i+1} = M\odot I_{org} +  (1-M)\odot I_{i}}
\end{split}
\end{equation}

Where $I_{org}$ is the original image, $M$ is the binary mask ($0$ for the masked region), $I_{i}$ is the output of the previous cycle and $I_{i+1}$ is the input of next cycle. $I_{0}$, which is the masked image, represents the input of the model in the first cycle. This iterative process is performed until a plausible image is obtained; more reasonable content is incrementally generated at each step in the masked area, after only a few iterations, i.e. usually less than 10.

\subsection{The Discriminator Network}
\label{sec:Discriminator_network}

\begin{figure*}
	\centering
	\includegraphics[width=.7\textwidth]{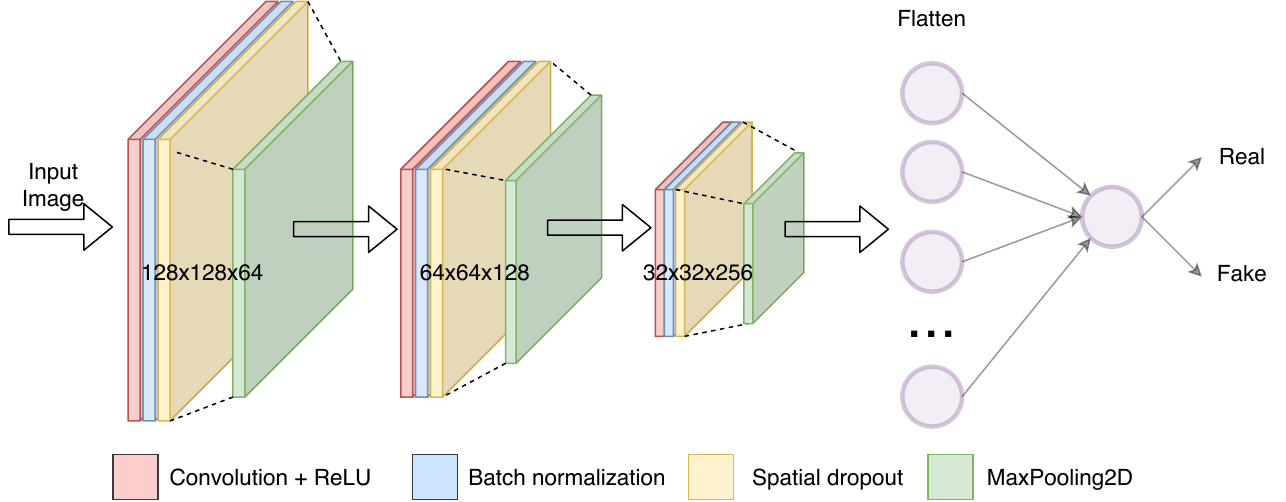}
	\caption{The discriminator network architecture.}
	\label{fig:discriminator}
\end{figure*}

In this section, we provide the details for the proposed discriminator network. In our method, it is important to determine which image looks visually realistic. For this purpose, we established a multi-layer convolutional neural network ending with a sigmoid unit. The structure of the discriminator is depicted in Fig. \ref{fig:discriminator}.

\begin{figure*}[!h]
	\centering
	\includegraphics[width=.6\textwidth]{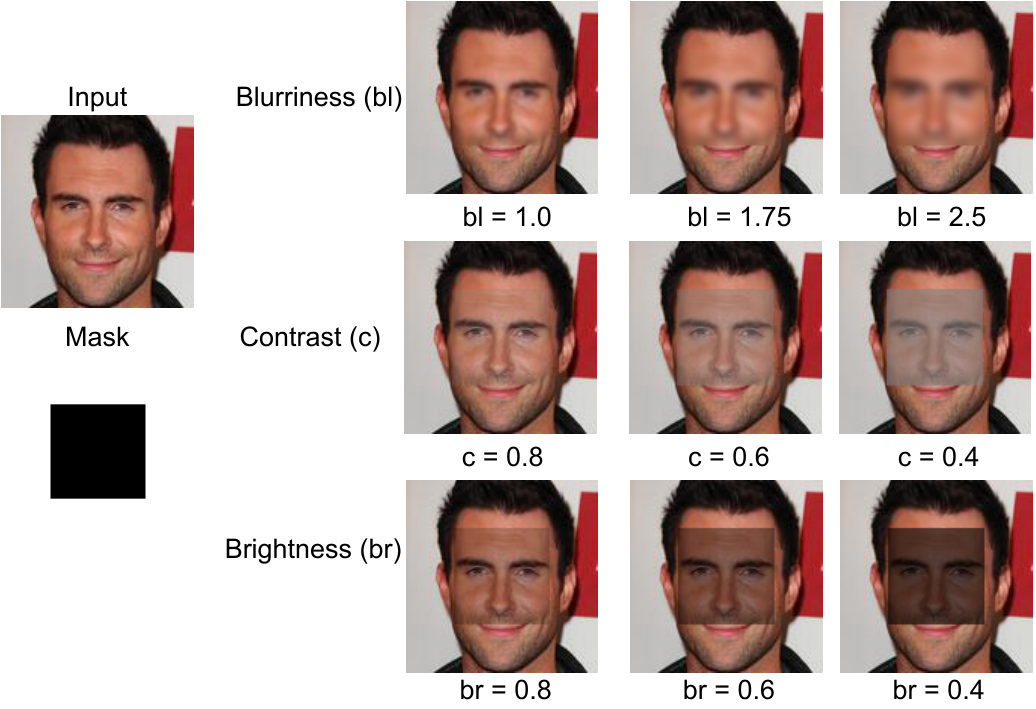}
	\caption{A sample image (on the left) with different scales of artifact adjustments performed on the masked region (on the right). Applied artifact values are provided below each image for blurriness, contrast and brightness adjustments.}
	\label{fig:artifact_range}
\end{figure*}

In general, a difference in brightness, blurriness, and contrast occurs between the masked region and the rest of the image in inpainting works. Our aim is to establish a model that will generate a score taking these differences into account. For this purpose, firstly, we created a dataset by generating regular and irregular masks and adding the visual artifacts we mentioned earlier to the masked regions to a certain amount. We adjusted the artifact range to ensure that the relevant artifact was applied at low, medium, and high levels. We applied artifacts by generating random values in the related ranges for blurriness $[1.0 - 2.5]$, brightness $[0.4 - 0.8]$, and contrast $[0.4 - 0.8]$, where a high value in blurriness and a low value in others increases the artifact ratio. In Fig. \ref{fig:artifact_range}, we provide samples taken from the left and right ends and middle of the ranges for the relevant artifact. Then, we labeled real images as 1 and images with the visual artifacts as 0. In the training phase, we used adversarial loss based on GAN, which is defined as:

\begin{equation}
	\label{equ:discriminator_equation}
	{\underset{D}{max}  \; {{\mathbb{E}} _{{x\sim p_{data}\left(x \right)}}} \log{\left(D\left(x \right) \right)}} \; + \;  { {{\mathbb{E}} _{{v\sim p_{VisArt}\left(v \right)}}} \log{\left(1- D\left(v \right) \right)}}
\end{equation}

Where $x$ and $v$ represent a real image and an image with visual artifacts, respectively.  We try to maximize the discriminator $D$ to distinguish real data, i.e. $p_{data}$, from the data with visual artifacts, i.e. $p_{VisArt}$. As the given task to the discriminator is simple, it converges quickly and produces only $0$ and $1$ scores for the input images, which is not useful for our purposes. We want to get gradual scores which reflect the visual quality of the images. Therefore, we changed our training policy to overcome this problem: First, we labeled the real images as $0.9$ (instead of $1$) and the distorted images as $0.1$ (instead of $0$). Second, we labeled some images that we applied very little artifacts by generating random values in the related ranges for blurriness $[0.3-0.8]$, brightness $[0.85-0.95]$, and contrast $[0.85-0.95]$ as real. Our dataset contains $60k$ images in total; $22k$ of these images are undistorted images and $8k$ of them are mildly distorted, both of which are labeled as real; and $30k$ images, which are highly distorted, are labeled as fake. We provide further details about the model architecture and training in the Supplementary Materials (Section \ref{supp_mats}).

\begin{figure}[!h]
	\centering
	\includegraphics[scale=.7]{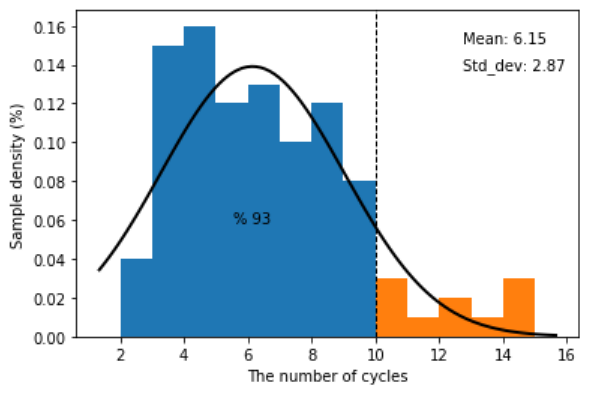}
	\centering
	\caption{Distribution of the number of samples with their maximum scores for a range of iterations.}
	\label{fig:hist_iteration_size}
\end{figure}

\subsection{Coarse Image Generation}
\begin{figure}[!h]
	\centering
	\includegraphics[scale=.4]{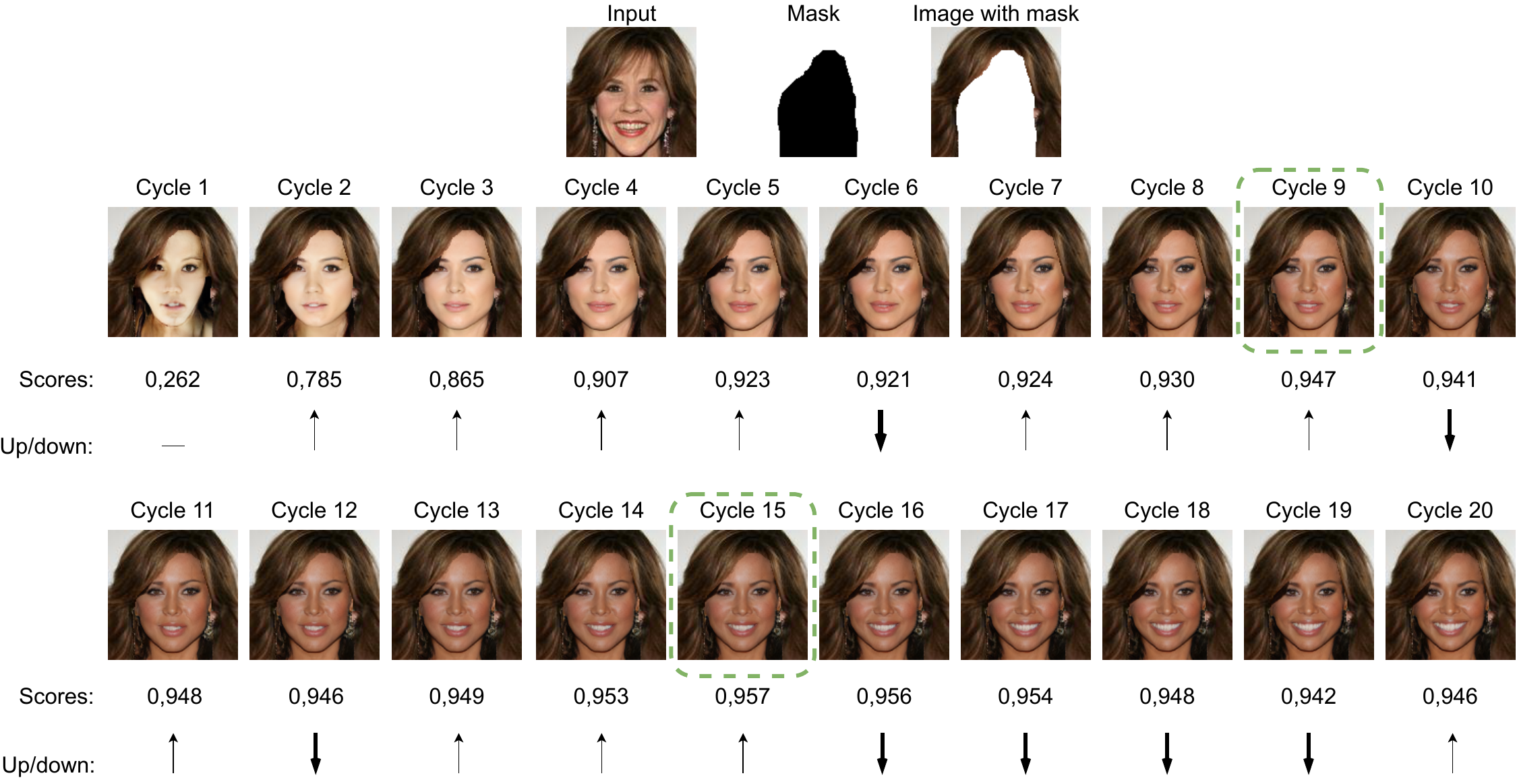}
	\centering
	\caption{A sample inpainting process is depicted with generated images with their discriminator scores for 20 iterations. Fluctuation of the scores are displayed visually with arrows below each cycle.}
	\label{fig:fluctuation_scores}
\end{figure}

An important design consideration of the proposed method is the determination of the convergence criteria for the images generated iteratively using our generator. For this purpose, we needed to evaluate the quality of the generated images in each iteration and stop when the image quality is sufficient for our coarse prediction stage. To automatize this process, we use our discriminator network that assesses the quality of the generated images and produce scores between $0$ and $1$, where score $1$ indicates high image quality (Section \ref{sec:Discriminator_network}). We provide the generated image at the end of each cycle to the discriminator and go on this process as long as image scores continue ascending i.e. image quality increases. As a precaution for small fluctuations in the scores, we configured our system for a predefined minimum number of iterations based on our empirical observations. To determine the feasible minimum iteration number, we randomly selected $100$ images and masks and generated the histogram of their evaluation scores for a varying number of iterations (Fig. \ref{fig:hist_iteration_size}). Note that, after $10$ iterations, we do not observe a significant increase in the scores of most of the images; for $93\%$ of the images, the highest image scores are obtained before $10$ iterations. For the remaining samples, as exemplified in Fig. \ref{fig:fluctuation_scores}, while the image with the highest score is obtained after the $10th$ iteration, the image with the highest score before $10th$ iteration, i.e. $9th$, is also visually plausible and sufficient for our coarse image generation task. This sample also exemplifies a small fluctuation in the evaluation scores before the $10th$ iteration, e.g. cycle $5$ and $6$, although the scores continue to increase afterward. Therefore, we set the minimum number of iterations to be 10 and choose the generated image with the highest score, if it stopped increasing somewhere in this interval.

Our proposed method has these advantages: (1) Arbitrarily structured masked regions can be used during the inference of our generator;  it is a generic framework that does not need any supervision depending on the masks, (2) Multiple and diverse results can be generated since masked image could be embedded to a different point in the generator latent space depending on the pixel values in the masked regions, and (3) Sketch-based inpainting is possible since the masked image with sketch would potentially be embedded to a different related point in the generator space, e.g. a mask put around the eye region with roughly sketched eyeglasses in the mask image, generates images with eyeglasses.

\begin{figure}[!h]
	\centering
	\includegraphics[width=0.7\textwidth]{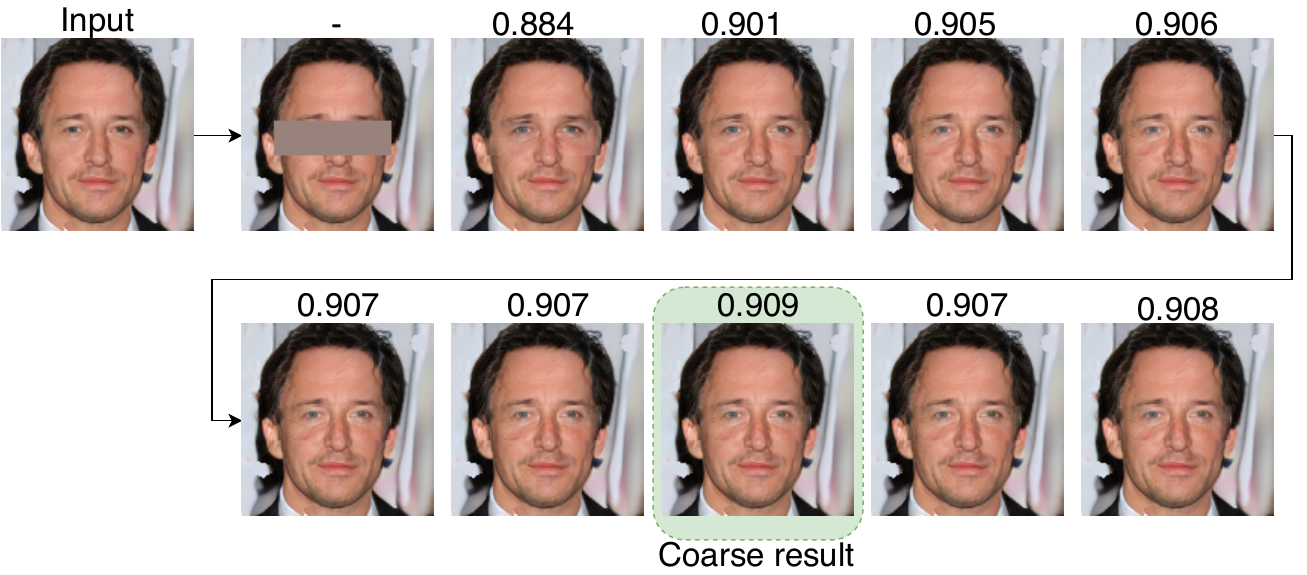}
	\centering
	\caption{Coarse result obtained using a mask containing the averages of the image pixel values.}
	\label{fig:average_value_coarse_result}
\end{figure}

In our method, depending on the initial pixel values assigned to the masked regions, an image can be embedded to a different point in the latent space, which is useful since this allows different alternative image generations. In Fig \ref{fig:average_value_coarse_result}, we masked the $128x128$ facial image to cover the eyes and assigned the average pixel values in the unmasked image region. Above each image, we provide the discriminator score of the image obtained at the end of that cycle. Scores vary between $0$ and $1$, and increase as the algorithm iterates using the generated image in the previous cycle. We iterate the CRG model until the stopping criterion is met and get the image with the highest score of $0.909$ in the $8th$ cycle as a coarse prediction. Note that the coarse result is without eyeglasses.

\begin{figure}[!h]
	\centering
	\includegraphics[width=0.7\textwidth]{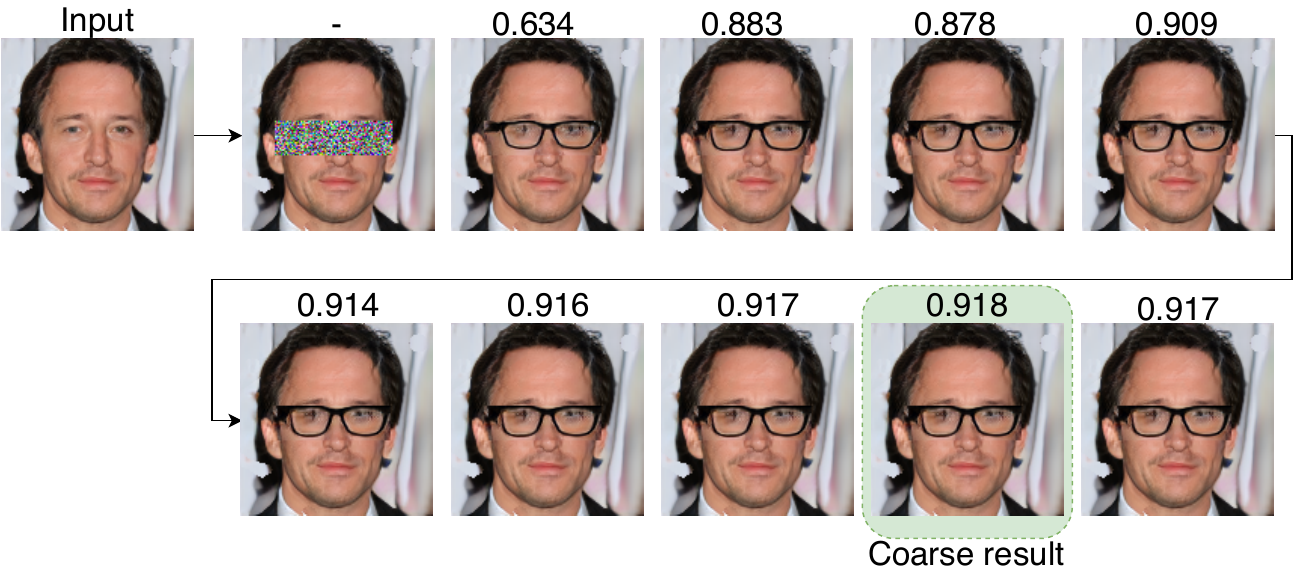}
	\centering
	\caption{Coarse result obtained using a mask containing random pixel values.}
	\label{fig:random_value_coarse_image}
\end{figure}

In the next example, we assigned random pixel values with zero mean to the masked region of the same face.  When the initial mask content is changed this way, the encoder embeds the masked image to a different latent point in the generator space, which results in different content at the end. In Fig. \ref{fig:random_value_coarse_image}, the masked region is filled with eyeglasses, differently from the final image shown in Fig. \ref{fig:average_value_coarse_result}. To the best of our knowledge, this feature in image inpainting is novel; almost all the methods either ignore the assigned pixel value in the masked area or use the same/fixed value assignment regime to the mask content that is used during the training phase.

\begin{figure}[!h]
	\centering
	\includegraphics[width=.7\textwidth]{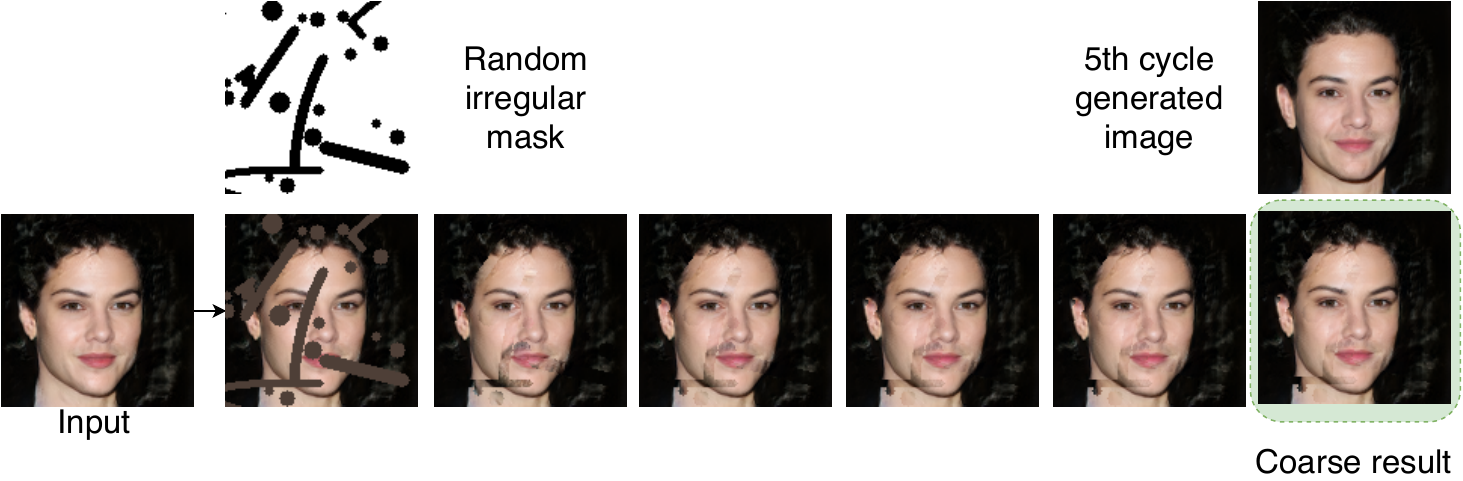}
	\caption{Coarse result obtained using a random irregular mask.}
	\label{fig:average_value_irregular_mask}
\end{figure}

In the proposed method, there is no restriction in the shape of the masks. As a showcase, we depicted a sample inpainting in Fig \ref{fig:average_value_irregular_mask} using an arbitrarily shaped, irregular mask for $5$ iterations. The resultant image is displayed in the upper right part of the Figure. Note that the generated image converges to the original image for this sample, and the masked content incrementally becomes more reasonable.

\subsection{The Unet Model}
\label{sec:unet_arch}
We need to preserve the structural continuity between a masked region and its surrounding context, for realistic-looking image inpainting. For this purpose, we train a Unet model, to eliminate visual artifacts in the converged coarse image around the mask boundaries. We adapted the Unet model from \cite{isola2017image, liu2018partial} and trained all its parameters from scratch using different loss functions. During the training phase, a mask and coarse CRG result are given to our Unet model as input and the original image is provided as the expected output. We aim to minimize the joint loss function in (\ref{equ:loss_joint}), which consists of  per-pixel reconstruction loss and  style loss \cite{gatys2016styleLoss} calculated for both $I_{Unet}$ and $I_{comp}$. Given the converged CRG result, $I_{CRG}$, binary mask, $M$, and the Unet prediction $I_{Unet}$;  $I_{comp}$ is defined in (\ref{equ:I_comp}).

\begin{equation}
\label{equ:I_comp}
{I_{comp} = M\odot I_{CRG} +  (1-M)\odot I_{Unet}}
\end{equation}

\begin{equation}
\label{equ:loss_joint}
{L_{joint} = L_{recons} + k(L_{style_{Unet}} + L_{style_{comp}})}
\end{equation}

Per-pixel reconstruction loss is defined in (\ref{equ:L_recons}), where $I_{org}$ and $N_{org}$ denote original image and the number of pixels in $I_{org}$, respectively.

\begin{equation}
\label{equ:L_recons}
{L_{recons} = \frac{1}{_{N_{org}}}\left\| I_{Unet} - I_{org} \right\|_{1} }
\end{equation}

Style loss is provided for  $I_{Unet}$ in (\ref{equ:style_IUnet}) and for $I_{comp}$ in (\ref{equ:style_Icomp}). We use layers $pool1$, $pool2$ and $pool3$ as in \cite{liu2018partial} in the pretrained VGG-16 network \cite{simonyan2014VGG16} to calculate style losses. 

\begin{equation}
\label{equ:style_IUnet}
{{L_{style_{Unet}} = \sum_{j = 0}^{J-1}\frac{1}{C_{j}C_{j}}\left\|   K_{j}\left( G_{ \phi_{j}^{I_{Unet}}} - G_{ \phi_{j}^{I_{org}}}  \right)  \right\|_{1} }}
\end{equation}

\begin{equation}
\label{equ:style_Icomp}
{{L_{style_{comp}} = \sum_{j = 0}^{J-1}\frac{1}{C_{j}C_{j}}\left\|   K_{j}\left( G_{ \phi_{j}^{I_{comp}}} - G_{ \phi_{j}^{I_{org}}} \right) \right\|_{1} }}
\end{equation}

In (\ref{equ:style_IUnet}) and (\ref{equ:style_Icomp}), $\phi_{j}^{I_{*}}$ corresponds to activation map of the $j^{th}$ layer (i.e. $pool1$, $pool2$ and $pool3$) for input $I_{*}$. The shape of $\phi_{j}^{I_{*}}$ is $C_{j}\times H_{j} \times W_{j}$ where $C$, $H$ and $W$ are the channel size, height and width of related layer, respectively. $G_{ \phi_{j}^{I_{*}}}$ is a $C_{j}\times C_{j}$ Gram matrix, i.e. covariances of the activation map of $\phi_{j}^{I_{*}}$, and $K_{j}$ is normalization coefficient, i.e. $1/_{C_{j} H_{j} W_{j}}$, for $jth$ activation layer.

\section{Experiments}
\label{sec:experiments_and_results}

\subsection{Elimination of Visual Artifacts}
\begin{figure}[!h]
	\centering
	\includegraphics[width=1.0\textwidth]{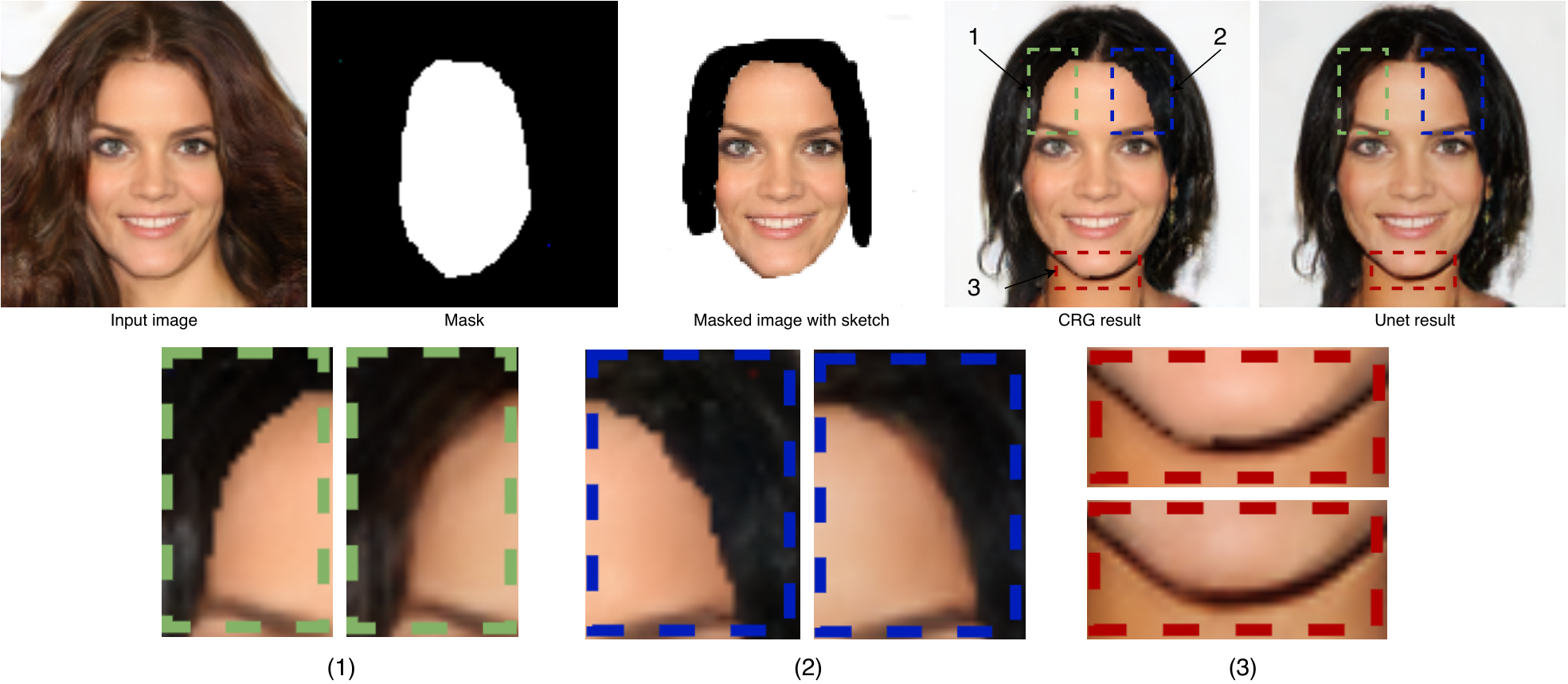}
	\caption{Elimination of visual boundary artifacts after post-processing. The parts in the green, blue and red parts are shown with $300\%$ zooming. (1)-left, (2)-left and (3)-top images are raw CRG outputs. (1)-right, (2)-right, (3)-bottom images are obtained after post-processing using Unet model.}
	\label{fig:removing_boundary_artifacts}
\end{figure}

In Fig. \ref{fig:removing_boundary_artifacts}, we show a sample removal of boundary artifacts with post-processing using our Unet model. In this sample, we first created a hand-drawn mask to keep the face part unchanged and applied the created mask to the image. We then roughly sketch black hair to the masked image to change the hair style to a darker one and use it as an input to our CRG model. After obtaining the inpainting result from the CRG model, we sampled three regions from both the coarse CRG result and Unet result for comparison. The improvement after post-processing in the boundary regions can be seen clearly in the image pairs in the bottom row; we observe better structural consistency and smooth color transition between the masked area and its surroundings after post-processing.

\paragraph{Effective range for the $k$ parameter:}

\begin{figure}[!h]
	\centering
	\includegraphics[width=1.0\textwidth]{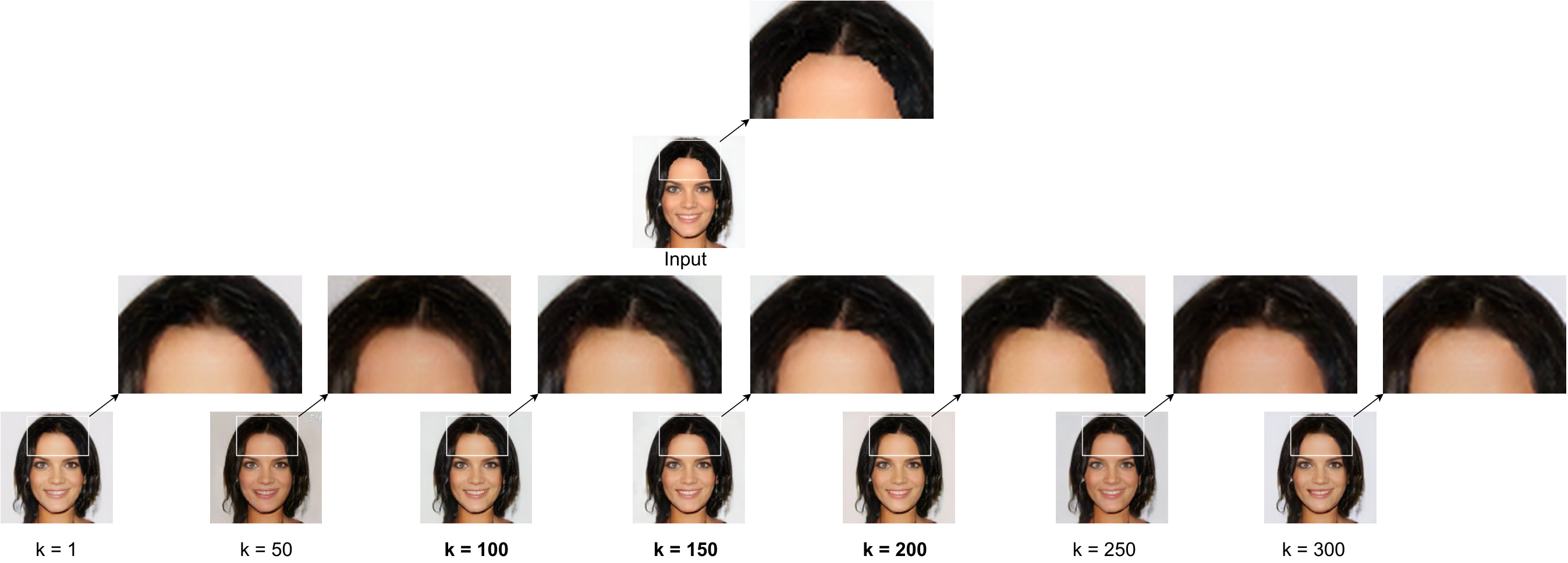}
	\caption{Effect of different $k$ values in Eq. (\ref{equ:loss_joint}) to the Unet results.}
	\label{fig:effective_range_for_k_parameter}
\end{figure}

We performed empirical evaluations to determine the effective value of the parameter $k$ in Equation (\ref{equ:loss_joint}). In this context, we retrained the Unet model from scratch for a selected set of $k$ values in [1, 300] intervals to observe the effect of different $k$ values. While adjusting $k$, we aimed to preserve the structural consistency with the generated coarse image in the first stage as much as possible, as well as the smooth color transition between the masked area and its surroundings. In Fig. \ref{fig:effective_range_for_k_parameter}, for each value of $k$, we cropped the parts covering the regions ($1$) and ($2$) depicted in the Fig. \ref{fig:removing_boundary_artifacts} and provided the Unet model results with $300\%$ zooming. We observed that structural continuity and color transition are provided for almost all $k$ values with minor differences; however, when we consider the details in the hairline pattern in that region, the content or style of the image changes dramatically when we use small, i.e. $k<100$, or high, e.g. $k>200$, values, respectively. Based on our empirical observation, for $k=[100-200]$ range, boundary artifacts of the coarse image are removed substantially and the other input features are kept very similar; hence, in our experiments we used the Unet model that we trained using $k=150$, i.e. the mean value in this range.

\subsection{ Capabilities of the Proposed Method}
In this section, we empirically examine and show the capabilities of our approach. Firstly, our method is advantageous for producing different alternative fillings in the mask region depending on the mask content. For instance, in Fig. \ref{fig:different_mask_different_pixel_value}, after masking the beard region in the facial image and filling it with white pixels in $(a)$, and black pixels in $(b)$, we observe that, depending on the initial mask content, our method generates different and semantically meaningful content in the masked region. The model generates a beardless image when the mask content is white and an image with beard when it is black. Secondly, our method allows the use of different mask types, i.e. hand-drawn (e.g., $(a)$, $(b)$, etc.), irregular, e.g., $(e)$, and regular masks (e.g., $(f)$, $(g)$) etc. Thirdly, our method enables sketch-based inpainting since the masked image with sketch embeds to a related latent vector position. For example, in $(g)$ we created a mask around the eye region with a thin-sketch similar roughly to eyeglasses, and as a result, a facial image with eyeglasses is produced; similarly, in $(h)$, we created a mask with a solid-sketch similar roughly to sunglasses, and consequently a facial image with sunglasses is produced. Finally, our method allows multiple and diverse results. To illustrate this, we provide the result of each cycle for $(k)$ in the right-bottom block. Note that several cycle results are useful; the discriminator could be disabled and each cycle result could be given to the Unet model as an input to get multiple results.

\begin{figure}[!h]
	\centering
	\includegraphics[width=1.0\textwidth]{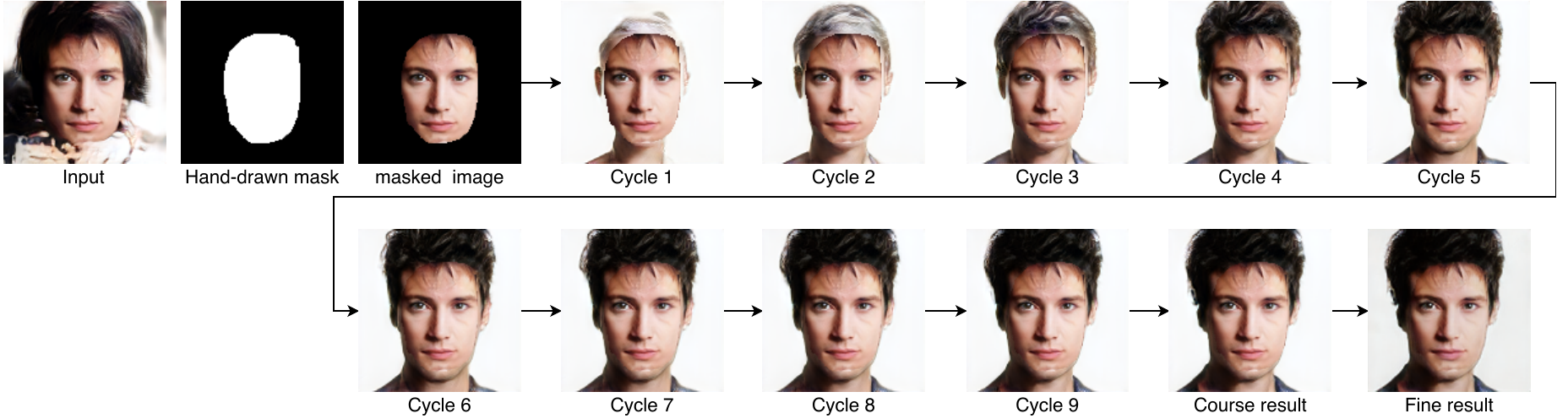}
	\caption{Inpainting unwanted regions.}
	\label{fig:generated_image_correct}
\end{figure}

Despite the advances in GANs, training a generator model is still challenging; two main problems arise in general \cite{dogan2019stability}. The first one is the stability problem; competition between the two networks, i.e. generator and discriminator, creates instability during training since one dominates the other. The second problem is the mode collapse problem; the generator collapses which produces a limited variety of samples. Recently, many tips and tricks \cite{salimans2016improved}, loss functions\cite{martin2017wasserstein, gulrajani2017wgan_gp}, or network architectures \cite{radford2015dcgan, karras2020stylegan2} have been proposed for dealing with these problems, but still it is an active area of research. For instance, the generator models may sometimes produce distorted images in some regions, i.e. images with acceptable face content but fail in the surroundings. Our method could also be useful to remedy the problems in such generated images by masking the parts that are distorted or we want to change or improve parts of those images. In Fig. \ref{fig:generated_image_correct}, we depict an example image that is generated randomly using the CRG generator. When the generated image is examined, it is seen that the face part is plausible but the surrounding context could be improved. We created a hand-drawn mask for that purpose, to keep the face part unchanged and enable reproduction of the surrounding regions. Note that as the number of cycles progresses, the masked part improves and becomes more coherent with the face. 
 
\subsection{Model Comparisons}
In this section, we compare our method with the state-of-the-art models designed for image inpainting, both quantitatively and qualitatively. Additional results with the real images with different mask types are also provided in the Supplementary Materials (Section \ref{supp_mats2}).

\begin{figure}[!h]
	\centering
	\includegraphics[width=1.0\textwidth]{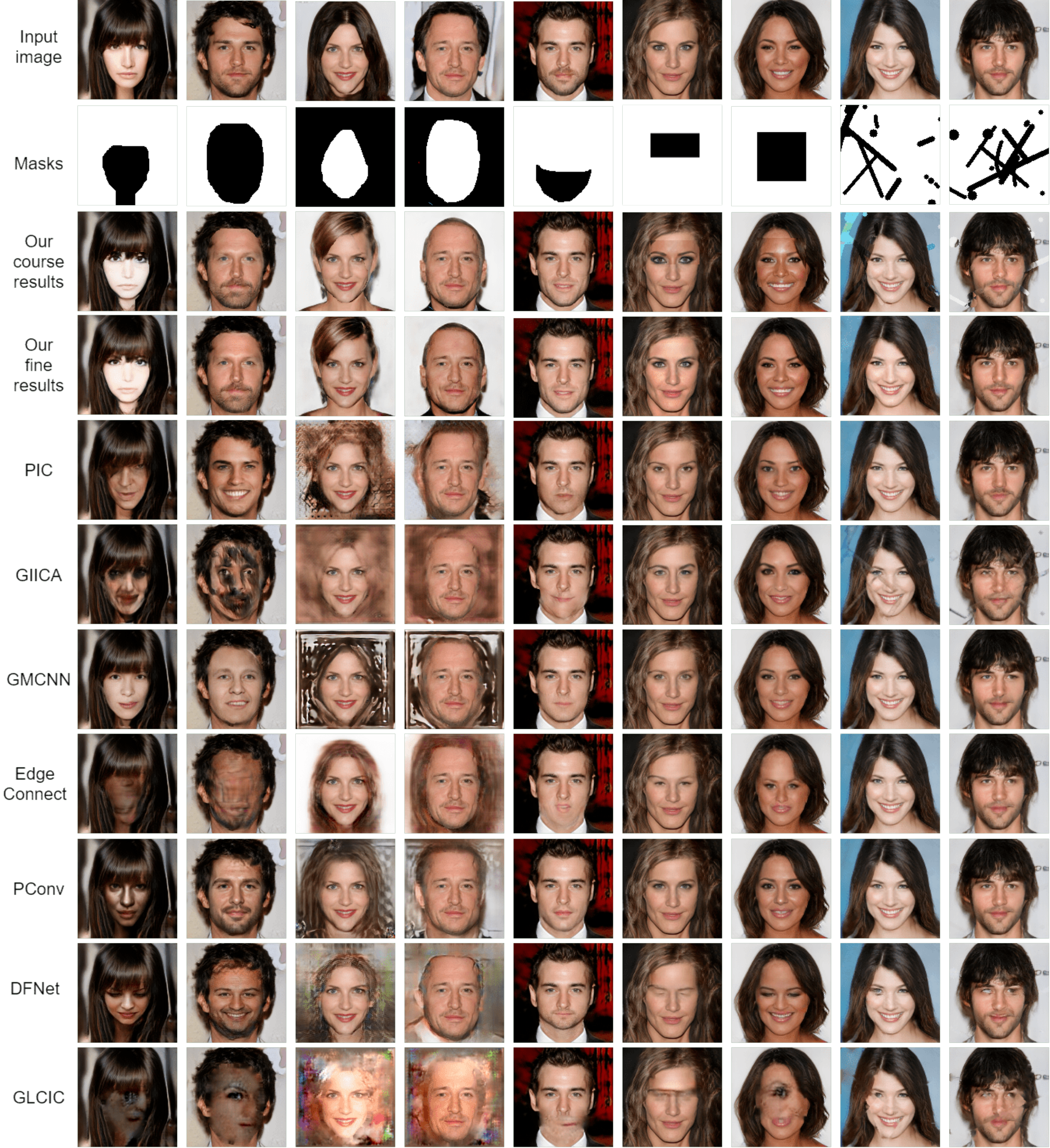}
	\caption{Comparison of the model results.}
	\label{fig:compare_models}
\end{figure}

\paragraph{Qualitative comparisons:}
In the vast literature, facial inpainting models have been trained using different mask types at different resolutions. In our method, we utilize the CRG model, which does not require any supervision depending on the mask types. Therefore, our method is flexible considering the mask types, sizes, and placements in the image. We compare our method performance qualitatively with the state-of-the-art models in the facial inpainting domain, i.e.
PIC \footnote{\url{https://github.com/lyndonzheng/Pluralistic-Inpainting}}\cite{zheng2019pluralistic}, 
GLCIC \footnote{\url{https://github.com/otenim/GLCIC-PyTorch}}\cite{lizuka2017globally}, 
GIICA \footnote{\url{https://github.com/JiahuiYu/generative_inpainting/tree/v1.0.0}}\cite{yu2018contextual}, GMCNN\footnote{\url{https://github.com/shepnerd/inpainting_gmcnn}}\cite{wang2018gmccn}, EdgeConnect\footnote{\url{https://github.com/knazeri/edge-connect}}\cite{nazeri2019edgeconnect}, 
DFNet\footnote{\url{https://github.com/hughplay/DFNet}}\cite{hong2019deepFusion}, 
Pconv \footnote{\url{https://github.com/MathiasGruber/PConv-Keras}}\cite{liu2018partial} for different mask types, i.e. hand-drawn $(1-5)$, regular $(6-7)$, and irregular $(8-9)$. Some of these works have trained their own models using regular masks, e.g., GIICA and GLCIC, while some of them, e.g., EdgeConnect, DFNet and Pconv, use irregular masks. There is also another group that supports both regular and irregular masks, e.g., PIC and GMCNN. During the experiments we obtained the results using official pre-trained models, if available; otherwise, we sent the masks and inputs to the model authors and get the results from them, used available pre-trained models that are configured as explained in the original articles, or retrained the models from scratch. We resize the original image and mask for models that require different resolutions. For the PIC model, there are two published pre-trained models depending on the mask placement; one serves when the mask is in the center of the image, the other serves for random placement. Similarly, the GMCNN model has two different pre-trained models for regular and irregular masks. We used the corresponding pre-trained model depending on the mask type. Also, the PIC model can generate multiple results for one mask, so we chose the best result based on their discriminator score. The comparison of the models using the generated images is depicted in Fig. \ref{fig:compare_models}. As can be seen from the results, our method can produce qualitatively good results in different mask types and produce better results particularly in the hand-drawn mask types (i.e, 1-5). Note that when the masked region gets larger, most of the models produce poor results. Our method does not use any masks in the training phase; thanks to the generator, it works also well in the case of larger masks.

\paragraph{Test Data:}
In order to provide quantitative analysis, we created a test set that contains $1000$ images that is not included in any stage in the training of our models\cite{jo2019sc}. Image inpainting works have been trained using different mask types and most of the models perform well with small localized masks. In order to challenge the models, we created canonical masks that cover a large region, which includes all facial landmarks, in a face. Note that, each mask in this setting is an irregular, hand-drawn-like mask that is uniquely generated for that face image automatically. We expect the models to inpaint primary facial features in the masked regions realistically, considering the integrity of its surrounding context, i.e. hairstyle/color, skin color, head orientation, and so on. 

\begin{figure*}[!h]
	\centering
	\includegraphics[width=.6\textwidth]{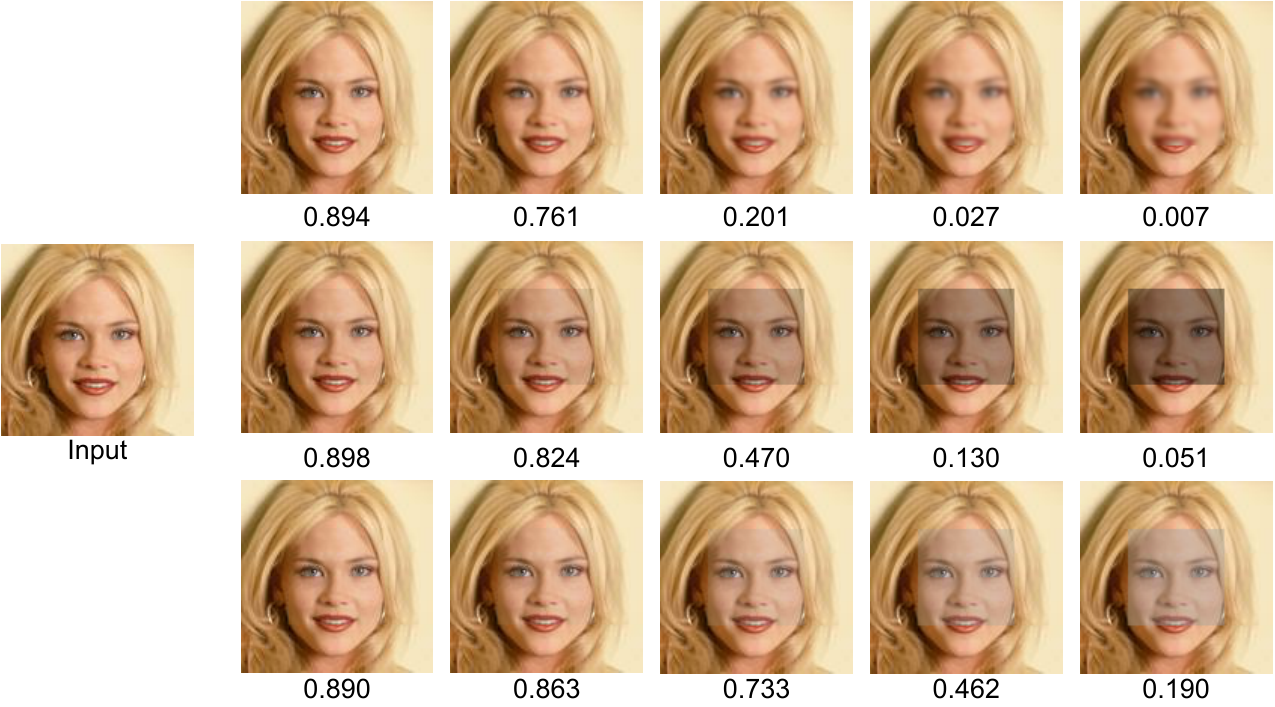}
	\caption{Application of three visual artifacts and image quality scores of our discriminator. Top row: bluriness, middle row: brightness and bottom row: contrast.}
	\label{fig:scoring_images_using_discriminator}
\end{figure*}

\paragraph{Quantitative comparisons:}

For quantitative comparisons, facial inpainting works report results using L1 error, PSNR, SSIM \cite{wang2004SSIM}, and Fréchet Inception Distance (FID) \cite{salimans2016improved} in general. However, as also mentioned in \cite{yeh2017semantic, yu2018contextual, wei2019facial}, these metrics are not exactly serving for good and fair evaluations of the methods; due to many possible solutions different from the original image content. These metrics generate scores based on the similarity of the produced results with the original image. However, this is not something necessarily desired in this domain, e.g., sample shown in Fig. \ref{fig:generated_image_correct}. In our work, we specifically designed and trained a discriminator network that generates scores depending on the visual quality of the given image. To show the consistency of its scores, we depicted a sample image in Fig. \ref{fig:scoring_images_using_discriminator} with three visual artifacts, i.e. blurriness, brightness and contrast, that are applied in the masked regions in an increasing amount. As can be seen in this Figure, as the amount of artifact increases, the scores decrease in a consistent manner. The images on the left-most column have the least amount of deformation and their quality scores are very close to each other. On the right-most column, as the contrast change (third row) does not damage the face as bad as blurriness and brightness artifacts, i.e. primary facial features are relatively sharper and more apparent, its score is relatively better than the other two. As these results and our previous samples in the coarse image generation process demonstrate, the discriminator scores are useful for the evaluation of visual qualities of the generated facial images. The discriminator learned the distribution of facial images, and it is also trained to be sensitive to the image's visual quality. Since it does not require reference images for comparisons and scoring, it is advantageous for overall performance assessments of the generated facial images. 
\begin{figure}[!h]
	\centering
	\includegraphics[width=1.0\textwidth]{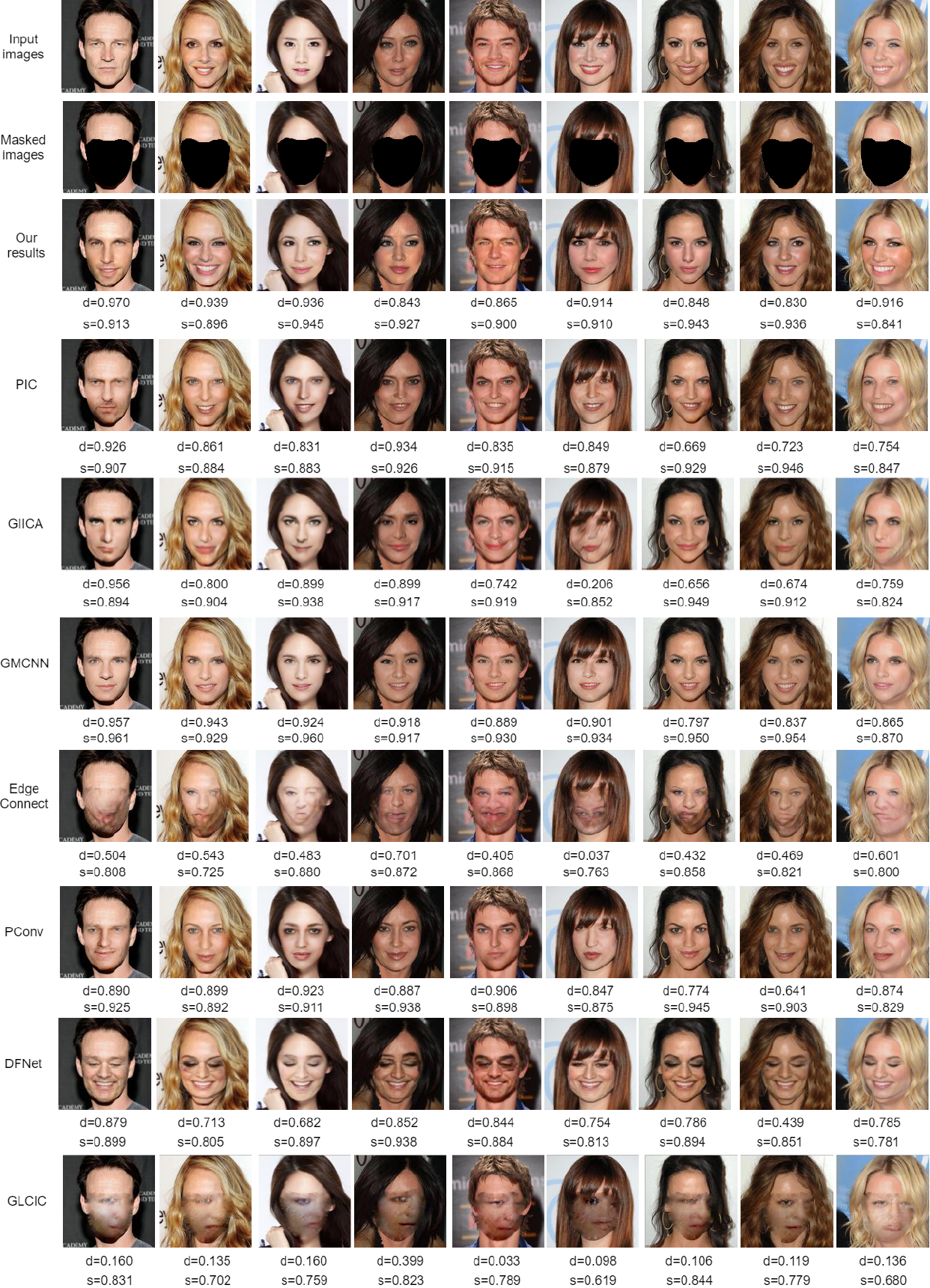}
	\caption{Comparison of models by masking all landmarks in facial images. Below each image, \textit{d} stands for the discriminator score, \textit{s} stands for SSIM scores.}
	\label{fig:masking_all_facial_features}
\end{figure}

We provided some samples, masked images, and generated images of each method from the test set in Fig. \ref{fig:masking_all_facial_features}. We included the visual quality scores (produced by our discriminator) and SSIM scores below each result in the same figure. The average quality scores of the models for the entire test set is provided in Table \ref{tab:table2}. As can be seen from the table, our overall visual quality score is better than the other models, i.e., $0.83$; yet very close to the overall scores of GMCNN and PConv models. These two models also generate realistic facial features. These scores are consistent with the qualitative results as can be seen from the Figure. The lowest average scores belong to GLCIC method; the visual samples (at the bottom row of Fig. \ref{fig:masking_all_facial_features}) show that the method fails to place facial features in the face correctly in this setting. Edge Connect method also has relatively lower average scores than the others; in the visual samples of their results, we observe highly deformed image features in the generated faces. The best overall FID score is obtained by PIC model and the best SSIM score is obtained by GMCNN method. Our method also has comparable FID and SSIM scores with these models. As we pointed out before, FID and SSIM computations require reference images. For those metric calculations, we provided input images as references for the assessments. Particularly, SSIM checks structural similarity using pairs of images, which is not particularly suitable for our purposes here. As can be seen under the sample images below our method in Fig. \ref{fig:masking_all_facial_features}, images look qualitatively good and realistic, yet some of the SSIM scores are lower than we expected, e.g. the right-most image, due to the structural differences with the input images. Similarly, we also observe  \textit{unfair} scoring patterns for the other methods' scores as well (e.g. right-most images of GMCNN, PConv, DFNet models).

\begin{table}[!h]
	\caption{Quantitative comparison of the models using CelebA dataset.}
	\centering
	\label{tab:table2}
		\begin{tabular}{|c|c|c|c|c|c|l|c|c|}
			\hline
			\textbf{Model} &
			\textbf{\begin{tabular}[c]{@{}c@{}}Our\\ Model\end{tabular}} &
			\begin{tabular}[c]{@{}c@{}}PIC\\ \cite{zheng2019pluralistic} \end{tabular}  &
			\begin{tabular}[c]{@{}c@{}}GIICA\\ \cite{yu2018contextual}\end{tabular}  &
			\begin{tabular}[c]{@{}c@{}}GMCNN\\ \cite{wang2018gmccn}\end{tabular}  &
			\begin{tabular}[c]{@{}c@{}}Edge\\ Connect\end{tabular} \cite{nazeri2019edgeconnect} &
			\begin{tabular}[c]{@{}c@{}}PConv\\ \cite{liu2018partial}\end{tabular}  &
			\begin{tabular}[c]{@{}c@{}}DFNet\\ \cite{hong2019deepFusion}\end{tabular}  &
			\begin{tabular}[c]{@{}c@{}}GLCIC\\ \cite{lizuka2017globally}\end{tabular}  \\ \hline
			\textbf{\begin{tabular}[c]{@{}c@{}}Visual \\ Quality\end{tabular}} &
			\textbf{0.83} & 0.75 & 0.69 & 0.82 & 0.40 &	0.82 &	 0.75 & 0.20 \\ 
			\hline
			\textbf{FID} &
			26.12 & \textbf{25.32} & 29.64 & 31.60 & 44.09 &	27.24 &	31.16 & 57.77 \\ 			
			\hline
			\textbf{SSIM} &
			0.87 & 0.87 & 0.88 & \textbf{0.90} & 0.79 & 0.86 &	0.84 & 0.76 \\ 			
			\hline
		\end{tabular}
\end{table}

\paragraph{Cross Dataset Inpainting Results:}

\begin{figure}[!h]
	\centering
	\includegraphics[width=1.0\textwidth]{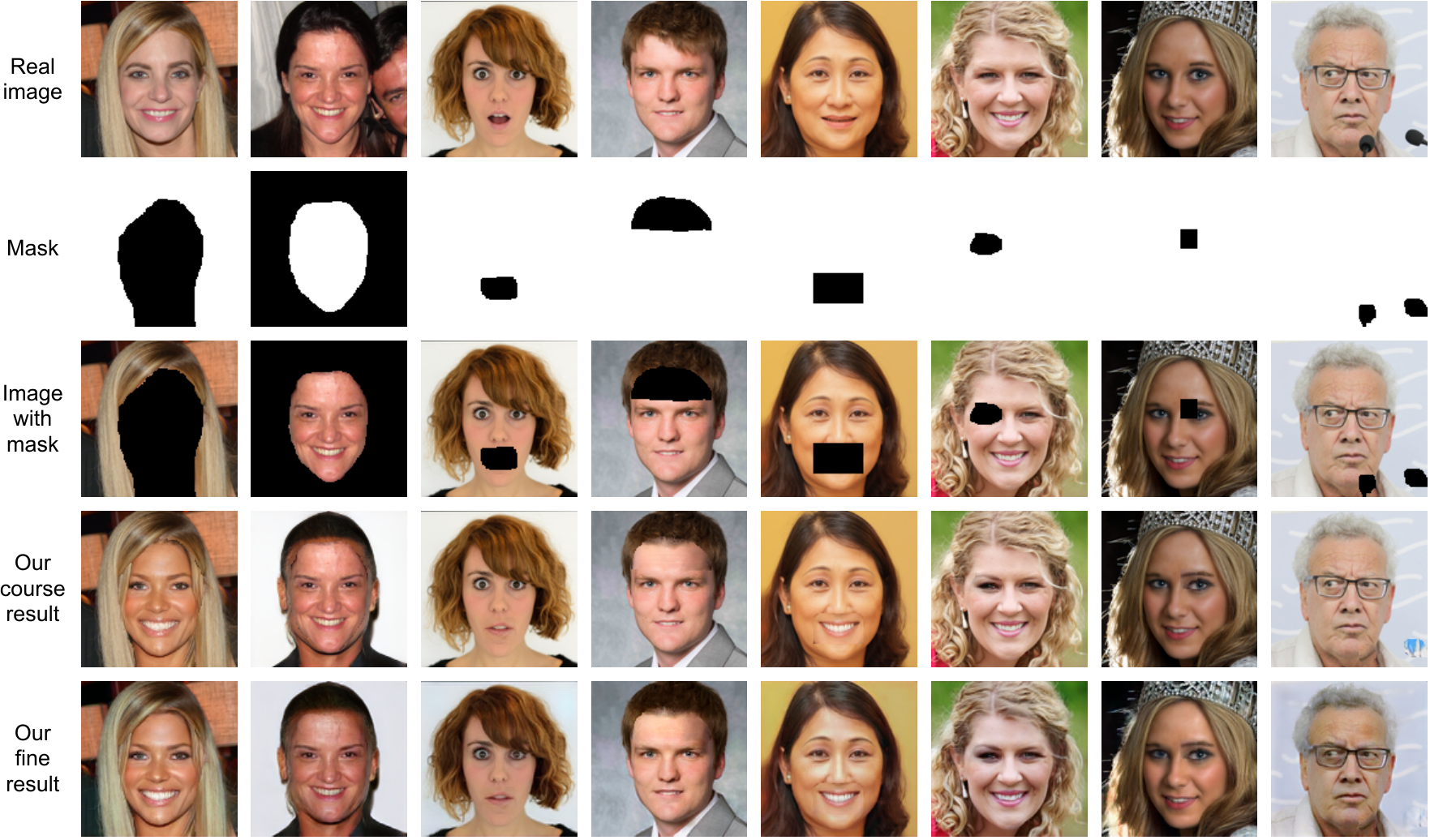}
	\caption{Cross dataset evaluation using FFHQ dataset.}
	\label{fig:real_images_ffhq}
\end{figure}

 The generator and encoder models that we use in this work (\cite{dogan2020semi}) are trained using $30K$ ($28K$ train, $1K$ validation, and $1K$ test set) facial images from CelebA dataset. In order to evaluate the performance of our method on a different dataset, we performed some additional empirical observations using Flickr-Faces-HQ (FFHQ) dataset \cite{karras2018style}. Sample results with different mask types are depicted in Fig. \ref{fig:real_images_ffhq}. Note that the results are similar to the ones we depicted from CelebA dataset. For quantitative analysis, we created a test set that contains 1000 random images from the FFHQ dataset which is not included in any stage of the training phase. Similar to CelebA dataset experiment setting, each mask is an irregular, hand-drawn-like mask that is uniquely generated for that face image automatically. We provided the average quality scores of the models for FFHQ dataset in Table \ref{tab:table3}. We observe that although our method performs comparable to the state-of-the-art methods, the average scores are not as good as some methods in the table. Since the generator is trained using CelebA dataset and the distributions of the FFHQ and CelebA datasets are different, FFHQ images are embedded in dead zones more frequently. We provide further discussion on this in the upcoming section.

\begin{table}[!h]
	\caption{Quantitative comparison of the models using FFHQ dataset.}
	\centering
	\label{tab:table3}
	\begin{tabular}{|c|c|c|c|c|c|l|c|c|}
		\hline
		\textbf{Model} &
		\textbf{\begin{tabular}[c]{@{}c@{}}Our\\ Model\end{tabular}} &
		\begin{tabular}[c]{@{}c@{}}PIC\\ \cite{zheng2019pluralistic} \end{tabular}  &
		\begin{tabular}[c]{@{}c@{}}GIICA\\ \cite{yu2018contextual}\end{tabular}  &
		\begin{tabular}[c]{@{}c@{}}GMCNN\\ \cite{wang2018gmccn}\end{tabular}  &
		\begin{tabular}[c]{@{}c@{}}Edge\\ Connect\end{tabular} \cite{nazeri2019edgeconnect} &
		\begin{tabular}[c]{@{}c@{}}PConv\\ \cite{liu2018partial}\end{tabular}  &
		\begin{tabular}[c]{@{}c@{}}DFNet\\ \cite{hong2019deepFusion}\end{tabular}  &
		\begin{tabular}[c]{@{}c@{}}GLCIC\\ \cite{lizuka2017globally}\end{tabular}  \\ \hline
		\textbf{\begin{tabular}[c]{@{}c@{}}Visual \\ Quality\end{tabular}} &
		0.77 & 0.73 & 0.60 & \textbf{0.81} & 0.29 &	0.69 &	 0.74 & 0.25 \\ 
		\hline
		\textbf{FID} &
		35.27 & \textbf{28.35} & 31.21 & 31.60 & 51.91 &	30.35 &	35.88 & 39.90 \\ 			
		\hline
		\textbf{SSIM} &
		0.80& 0.83 & 0.81 & \textbf{0.84} & 0.72 & 0.81 &	0.79 & 0.73 \\ 			
		\hline
	\end{tabular}
\end{table}

\subsection{Failure Cases}  
The sufficiency of the generator latent space is crucial for the success of our method; for a sufficient generator model, for almost every point from the prior z distribution, sampling results, i.e. generated images, contain plausible semantic content. In this context, it is important that masked image could be embedded in a region with sufficient variance in the generator latent space. In some locations in the latent space, however, the manifold of trained examples does not have sufficient density. These regions are referred to as dead zones \cite{white2016sampling}. Sampling from such regions will not be useful to fill the mask content properly for the next cycle due to the lack of variance around that point in the latent space. 

\begin{figure}[!h]
	\centering
	\includegraphics[width=1.0\textwidth]{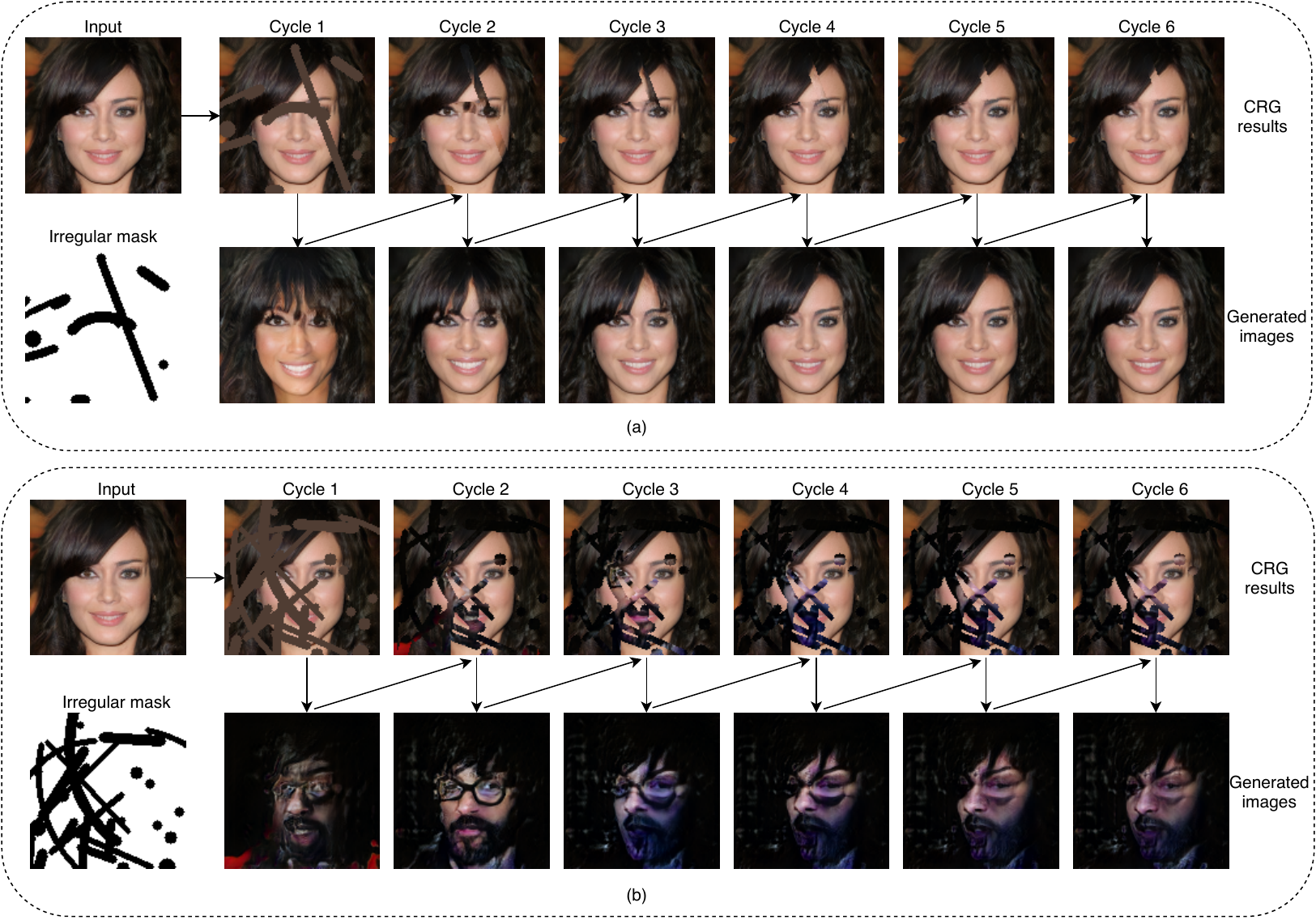}
	\caption{A sample for a failure case. For (a) and (b), top-row: shows the result of each the CRG cycle, bottom row: produced image in the generator space for the relevant cycle.}
	\label{fig:weakness_of_our_model}
\end{figure}

The level of defect applied to an image reinforces the probability of sampling from these dead zones, due to insufficient semantic content. An example is depicted in Fig. \ref{fig:weakness_of_our_model}; in $(a)$, the defect created to the semantic content of the input image with the applied mask is not as severe as the one depicted in $(b)$. The mask in $(b)$, hides important semantic content in the image, and makes embedding to the latent space more challenging for the CRG encoder. As can be seen from the second rows of both derivations, embedding the masked image in the first one, $(a)$, could be done in a region with sufficient variance, while for the second one, $(b)$, embedding is done in a dead zone. As can be seen in the generated images of $(b)$, the density around the embedded point is not high to recover a plausible face image in that area.

 \section{Conclusion and Future Works}
\label{sec:conclusion_and_future_works}
We propose an iterative algorithm to fill a masked region semantically with very few iterations via encoder-generator architecture. The encoder embeds a masked image to a latent point in generator space; since the content of the masked region directly affects the embedded point,  a variety of alternative fillings are generated depending on different drawn sketches or different pixel values assigned to the masked region. Also, since our solution uses generic frameworks, i.e. CRG, which does not need any supervision depending on the masks, it allows the use of different mask types. The empirical results show that our heuristic works successfully and generates comparable results with the state-of-the-art models both quantitatively and qualitatively, especially in images where the face or background is completely masked.

In time, the increase in the data and advanced computational resources will have a fueling effect in the progress in the deep generative models, which will make utilization of their rich representation capability more easy and convenient via integration of such models in different domains that require creative content generation (as a whole or in part as it is depicted here). This work provides general insight on how to utilize generative models in a creative content generation without direct intrusion into the generation process. The content is encoded in the generator latent space and we indirectly use it in a modular framework very efficiently. As more encoders for the generator models are trained, which is still a challenge in the domain, many alternative extensions through different data distributions will be readily available for use with the proposed method. It is important to note that, the success of our method depends on the sufficiency of the generator space and how well its inversion is performed; so, as future work, we are planning to make a more comprehensive analysis on the manifold of trained examples in the latent space utilizing the encoder of the CRG model.

\section*{Acknowledgements}
	This research is funded by Ankara University (Scientific Research Projects Grant, grant id: 18L0443010). The numerical calculations reported in this paper were partially performed at TUBITAK ULAKBIM, High Performance and Grid Computing Center (TRUBA resources). We would like to thank the anonymous reviewers for their valuable suggestions and comments.

\section*{Compliance with Ethical Standards}
\textbf{Conflict of interests} The authors declare that they have no known competing financial interests or personal relationships that could have appeared to influence the work reported in this paper.

\bibliographystyle{ieeetr}
\footnotesize \bibliography{Dogan_Keles_Inpainting}
\clearpage
\appendix
\section{The Architectures and Training Details}
\label{supp_mats}
\textbf{Unet architecture and training details:}
We adapt our Unet architecture from those \cite{isola2017image, liu2018partial}, replacing all partial convolutional layers with standart convolutional layers.  We only use masks to calculate style losses, so we do not follow the mask update steps. Table \ref{tab:table1} provides other architectural details. Each convolution layer is followed by a batch normalization layer. Similar to \cite{liu2018partial}, we perform down-sampling and up-sampling with a factor of 2,  we use Relu activation function in all convolution layers in the encoder, and LeakyRelu, with slope 0.2, in the decoder.

\begin{table}[!h]
	\caption{Unet architecture}
	\centering
	\label{tab:table1}
	\begin{tabular}{|c|c|c|}
		\hline
		\multicolumn{3}{|c|}{Encoder}                                                                                                                                                                        \\ \hline
		Layer name                                                                       & Filter size                                         & Number of filters                                           \\ \hline
		Conv1                                                                            & 7x7                                                 & 64                                                          \\ \hline
		Conv2                                                                            & 5x5                                                 & 128                                                         \\ \hline
		Conv3                                                                            & 5x5                                                 & 256                                                         \\ \hline
		Conv4                                                                            & 3x3                                                 & 512                                                         \\ \hline
		Conv5                                                                            & 3x3                                                 & 512                                                         \\ \hline
		Conv6                                                                            & 3x3                                                 & 512                                                         \\ \hline
		\multicolumn{3}{|c|}{Decoder}                                                                                                                                                                        \\ \hline
		\begin{tabular}[c]{@{}c@{}}Upsample1\\ Concat (with Conv6)\\ Conv7\end{tabular}  & \begin{tabular}[c]{@{}c@{}}-\\ -\\ 3x3\end{tabular} & \begin{tabular}[c]{@{}c@{}}512\\ 512+512\\ 512\end{tabular} \\ \hline
		\begin{tabular}[c]{@{}c@{}}Upsample2\\ Concat (with Conv5)\\ Conv8\end{tabular}  & \begin{tabular}[c]{@{}c@{}}-\\ -\\ 3x3\end{tabular} & \begin{tabular}[c]{@{}c@{}}512\\ 512+512\\ 512\end{tabular} \\ \hline
		\begin{tabular}[c]{@{}c@{}}Upsample3\\ Concat (with Conv4)\\ Conv9\end{tabular}  & \begin{tabular}[c]{@{}c@{}}-\\ -\\ 3x3\end{tabular} & \begin{tabular}[c]{@{}c@{}}512\\ 512+512\\ 512\end{tabular} \\ \hline
		\begin{tabular}[c]{@{}c@{}}Upsample4\\ Concat (with Conv3)\\ Conv10\end{tabular} & \begin{tabular}[c]{@{}c@{}}-\\ -\\ 3x3\end{tabular} & \begin{tabular}[c]{@{}c@{}}512\\ 512+256\\ 256\end{tabular} \\ \hline
		\begin{tabular}[c]{@{}c@{}}Upsample5\\ Concat (with Conv2)\\ Conv11\end{tabular} & \begin{tabular}[c]{@{}c@{}}-\\ -\\ 3x3\end{tabular} & \begin{tabular}[c]{@{}c@{}}256\\ 256+128\\ 128\end{tabular} \\ \hline
		\begin{tabular}[c]{@{}c@{}}Upsample6\\ Concat (with Conv1)\\ Conv12\end{tabular} & \begin{tabular}[c]{@{}c@{}}-\\ -\\ 3x3\end{tabular} & \begin{tabular}[c]{@{}c@{}}128\\ 128+64\\ 64\end{tabular}   \\ \hline
		\begin{tabular}[c]{@{}c@{}}Upsample7\\ Concat (with Input)\\ Conv13\end{tabular} & \begin{tabular}[c]{@{}c@{}}-\\ -\\ 3x3\end{tabular} & \begin{tabular}[c]{@{}c@{}}64\\ 64+3\\ 3\end{tabular}       \\ \hline
	\end{tabular}
\end{table}

We train the Unet model from scratch for 100 epochs, using Adam optimizer \cite{kingma2014adam} with a batch size of 16 and a small learning rate, i.e. 0.0002. To avoid overfitting or positional bias, we apply data augmentation where we shift the images left, right, up, or down and apply horizontal and vertical flips.

\textbf{The Discriminator network and training details:} We designed our discriminator as a multi-layer convolutional neural network depicted in Fig. \ref{fig:discriminator}. The encoder consists of 3  blocks and a fully connected layer with a total of approximately 437k trainable parameters. We use 64 filters in the first block and double the number of filters in consecutive blocks. We use a 3x3 filter size in the convolution layers. To avoid overfitting, we apply $50\%$ spatial dropout. We train the discriminator from scratch for 300 epochs, using RMSProp optimizer \cite{tieleman2012lecture} with a batch size of 128, setting learning rate to 0.0001, rho to 0.9 and epsilon to 1e - 08. We reduce the learning rate by a factor of 2 when minimum validation loss stops improving for 15 epochs and saves the best model using model checkpoint monitoring validation loss.

\section{Additional Qualitative Results}
\label{supp_mats2}

\begin{figure}[!h]
	\centering
	\includegraphics[width=1.0\textwidth]{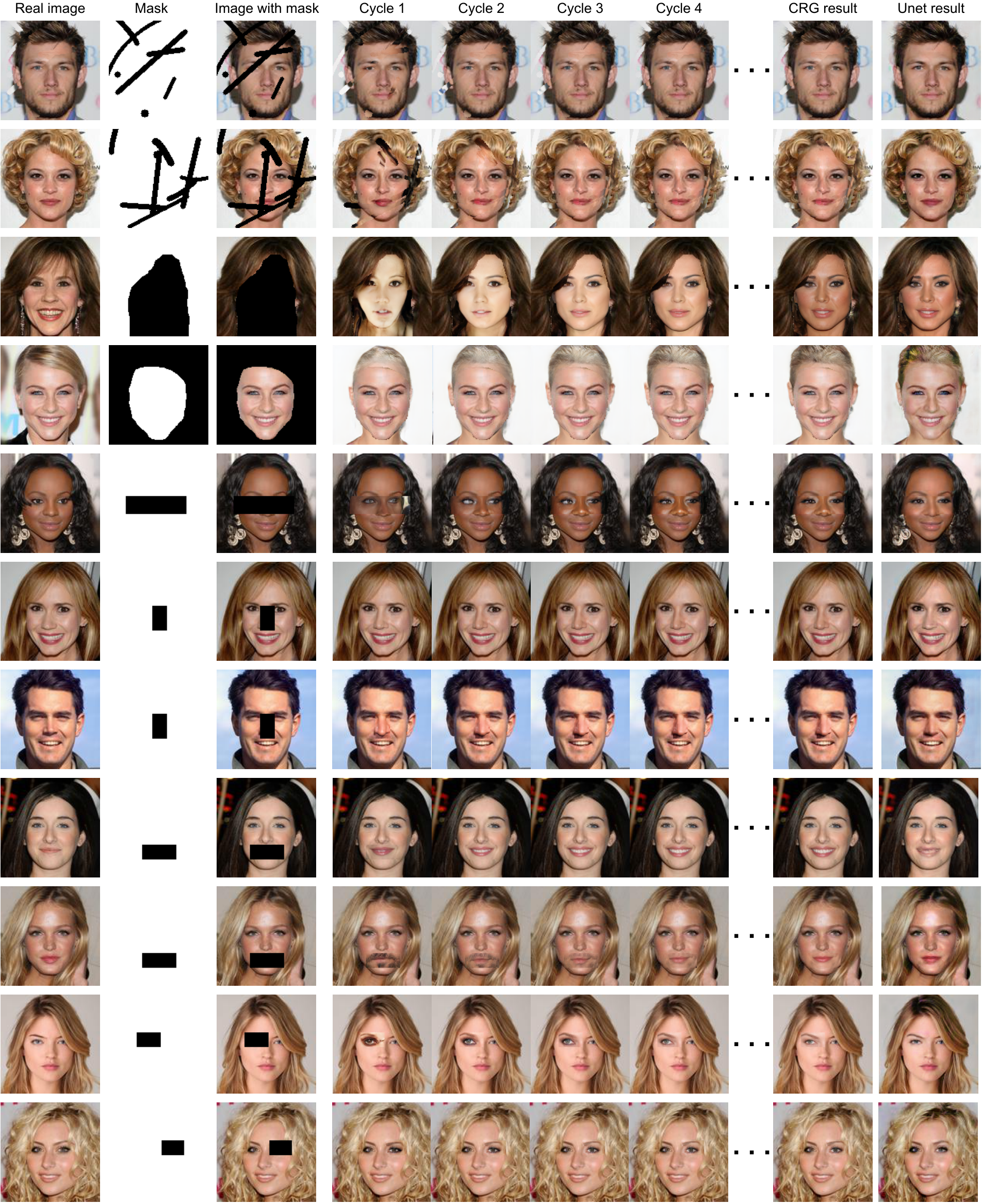}
	\caption{Sample inpainting with real images using a variety of different masks. }
	\label{fig:real_images}
\end{figure}

Additional sample results are provided in Fig. \ref{fig:real_images}. In some cases, our coarse CRG images visually look better than the post-processed Unet results, e.g. from top to bottom: 4 (color in the hair), 8 (in the mouth), and 11 (darkness in both eyes). Post-processing stage can be omitted for such cases, if desired, by either manual inspection or automatically utilizing discriminator scores.

\end{document}